\documentclass[12pt,notitlepage]{article}
\usepackage{draft}
\pdfoutput=1

\usepackage{latexsym,ifthen,lscape,epsfig,graphics,natbib,amsmath,amsfonts,amssymb,listings,mdframed}
\usepackage[margin=1in]{geometry}
\usepackage[hang,footnotesize,bf]{caption}
\usepackage{subfig}
\usepackage{color,array}
\definecolor{darkgreen}{rgb}{0,0.5,0}
\definecolor{darkblue}{rgb}{0,0,0.8}
\definecolor{expCol}{rgb}{0.6980392156862745, 0.29411764705882354, 0.23529411764705882}
\definecolor{msmdCol}{rgb}{0.8901960784313725, 0.6941176470588235, 0.37254901960784315}
\definecolor{tmsmdCol}{rgb}{0.4, 0.4392156862745098, 0.29411764705882354}
\definecolor{passiveCol}{rgb}{0.5058823529411764, 0.7333333333333333, 0.9764705882352941}
\usepackage{hyperref}
\usepackage{tabu}
\hypersetup{colorlinks, urlcolor=darkblue, citecolor=darkgreen}

\def\E{\mathbb{E}\,}
\def\Var{\mbox{Var}\,}

\def\eps{\varepsilon}

\makeatletter
\def\zapcolorreset{\let\reset@color\relax\ignorespaces}
\def\colorrows#1{\noalign{\aftergroup\zapcolorreset#1}\ignorespaces}
\makeatother

\begin{document}
\title{{\bf The Random Walk of High Frequency Trading}\thanks{This
work was supported by the Hellman Fellows Fund.}}
\author{
  Eric M. Aldrich\thanks{
  Email: ealdrich@ucsc.com.} \\
  \normalsize{Department of Economics} \\
  \normalsize{University of California, Santa Cruz}
\and
  Indra Heckenbach\thanks{
  Email: iheckenb@ucsc.com} \\
  \normalsize{Department of Physics} \\
  \normalsize{University of California, Santa Cruz}
\and
  Gregory Laughlin\thanks{
  Email: glaughli@ucsc.edu} \\
  \normalsize{Department of Astronomy and Astrophysics} \\
  \normalsize{University of California, Santa Cruz}
}
\date{First Draft: August 15, 2014 \\
      This Draft: \today}

\renewcommand{\baselinestretch}{1}
\maketitle
%
\renewcommand{\baselinestretch}{1}
\selectfont
\begin{center}
{\bf Abstract}
\end{center}
\par
\begin{footnotesize}
  \begin{quote}
    This paper builds a model of high-frequency equity returns by
    separately modeling the dynamics of trade-time returns and trade
    arrivals. Our main contributions are threefold. First, we
    characterize the distributional behavior of high-frequency asset
    returns both in ordinary  clock time and in trade time. We show
    that when controlling for pre-scheduled market news events,
    trade-time returns of the highly liquid near-month E-mini S\&P 500
    futures contract are well characterized by a Gaussian distribution
    at very fine time scales. Second, we develop a structured and
    parsimonious model of clock-time returns by subordinating a
    trade-time Gaussian distribution with a trade arrival process that
    is associated with a modified Markov-Switching Multifractal
    Duration (MSMD) model. This model provides an excellent
    characterization of high-frequency inter-trade
    durations. Over-dispersion in this distribution of inter-trade
    durations leads to leptokurtosis and volatility clustering in
    clock-time returns, even when trade-time returns are
    Gaussian. Finally, we use our model to extrapolate the empirical
    relationship between trade rate and volatility in an effort to
    understand conditions of market failure. Our model suggests that
    the 1,200 km physical separation of financial markets in Chicago
    and New York/New Jersey provides a natural ceiling on systemic
    volatility and may contribute to market stability during periods
    of extremely heavy trading.
    \\[2ex] 
  \end{quote}
\end{footnotesize}
\newpage

\renewcommand{\baselinestretch}{1.5}
\selectfont
\section{Introduction}
\label{intro}

Modern electronic exchanges function in a manner that outwardly
display the properties that are expected of an efficient, liquid
market. Bid-offer spreads are narrow in comparison to the price of the
underlying instrument being traded, volumes are high, high-frequency
traders compete to make markets, and information regarding price
discovery is disseminated at nearly the speed of light (see,
e.g. \citet{Brogaard2013} and \citet{Hasbrouck2013}). Within such an
environment, the disparate, shifting spectrum of intentions of a wide
range of market participants is continuously being aggregated, and so
a naive, but nonetheless reasonable, expectation is that the Central
Limit Theorem should play a fundamental role, and  that short-period
returns should adhere to a Gaussian distribution.

Indeed, from the pioneering work of \citet{Bachelier1900} through
the development of the Black-Scholes options pricing model
(\citet{Black1973}), modern finance has traditionally held that market
price movements can be approximated to a somewhat useful degree by a
Gaussian random walk. This concept draws reinforcement from the
Efficient Market Hypothesis (\citet{Fama1965}), since the arrival of
news shifts the market in an unpredictable manner
(\citet{Hasbrouck2003}).

In reality, observed distributions of market returns are
markedly non-Gaussian. Regardless of venue and asset class, returns
distributions invariably have fat tails, and display the
phenomenon of volatility clustering. A rich literature exists which
describes both the characterization and the modeling of the observed
departures from normality (for a review, see \citet{Bouchaud2005}).

In this contribution, we carry out a ground-level re-examination of
the process that generates short-period market returns within the
context of high-frequency trading (over time scales ranging from
milliseconds to minutes). We analyze three complete months of recent,
millisecond-resolution  tick data from the extremely liquid near-month
E-mini S\&P 500 futures contract traded at the Chicago Mercantile
Exchange.

Our analysis begins with two basic empirical conclusions. First, we
find that asset returns are effectively organized into two groups with
distinct properties: those that are the result of trade within a 1,000
second window following pre-scheduled news announcements (referred to
as the ``active'' period), and those occurring at all other times
(referred to as the ``passive'' period). Our analysis shows that
conditional on this sorting mechanism, returns distributions exhibit
very different tail characteristics as well as patterns of volatility
persistence. We suggest that this is attributable to the shocks of
pre-scheduled news announcements entering the market.

Second, we demonstrate that high-frequency returns in passive-market
periods are well described by a Gaussian distribution when trade time
is employed. \citet{Brada1966} introduced the notion of trade time to
show that asset returns distributions are nearly Gaussian if the
returns process is subordinated with successive transactions (trades)
acting as the subordinator. \citet{Mandelbrot1967} showed that a
Gaussian random walk composed with a subordinating trade-time process
is fully consistent with a fat-tailed, L\'{e}vy-stable distribution,
as suggested in \citet{Mandelbrot1963}. \citet{Clark1973} used an
alternative subordinator, time measured by volume of transactions, to
obtain similar results. More recently, \citet{Ane2000} show that
coarsely sampled intra-day returns also conform to a Gaussian
distribution when measured in trade (transaction) time.

Our analysis demonstrates that the Gaussianity of trade-time returns
does not immediately extend to high-frequency intra-day returns. That
is, high-frequency, trade-time returns exhibit heavy tails and
volatility clustering when considered unconditionally throughout the
day. However, by excluding the periods surrounding pre-scheduled news
events, we confirm the existence of trade-time Gaussianity as well as
a lack of volatility persistence.

Building on the empirical observations above, our paper makes two
theoretical contributions. First, we develop a model for
high-frequency inter-trade durations by expanding the Markov-Switching
Multifractal Duration (MSMD) model of \citet{Chen2013}. The MSMD model
builds on the seminal work of \citet{Mandelbrot1997},
\citet{Calvet1997} and \citet{Fisher1997}, as well as subsequent work
by \citet{Calvet2001}, \citet{Calvet2002} and \citet{Calvet2004}. It
suggests a parsimonious model of inter-trade durations that has a
well-founded physical interpretation and which has been shown to fit
historical data well. Since the MSMD model is unable to explain
certain features of high-frequency inter-trade durations, we modify
the model by composing an MSMD process with waiting times drawn from
an Exponential distribution that is chosen so that its maximal
expected duration matches the maximal value observed in our
sample. Intuitively, this process describes an environment with two
types of traders: the majority type, which typically trades quickly
and with infrequent pauses (e.g. algorithmic traders), and a minority
type which places few orders, spaced by more moderate durations
(institutional investors). The resulting Truncated Markov-Switching
Multifractal Duration (TMSMD) model does a better job of fitting the
empirical density of high-frequency inter-trade durations as well as
capturing the strong autocorrelation of volatility.

The second theoretical innovation of our paper is that we couple a
trade-time Gaussian random walk with the TMSMD process above to
characterize the distribution of asset returns in clock time. In
particular, we demonstrate that the subordinating transformation
between clock time and trade time for our data can be effectively
explained using the the TMSMD model. In this dimension
our work differs substantially from that of \citet{Ane2000}: where
they begin with a nonparametric estimate of the distribution of
clock-time asset returns and work backwards to implicitly define the
nonparametric density of trades that would be consistent with
trade-time Gaussianity, we work forwards by first compounding a
parametric distribution of trade-time returns with a parametric model
of duration times (and hence, an associated trade arrival process) to
characterize the distribution of clock-time returns. Our contribution
is significant because it promotes a structured and parsimonious
approach to approximating the observed evolution of asset returns.

Finally, we use our hierarchical model of clock-time returns to fit
and extrapolate the empirical relationship between inter-trade
duration (alternatively, trade rate) and market volatility. Our model
is suggestive of conditions under which trade rate could induce
unprecedentedly high levels of volatility and potential market
failure. We highlight how this could occur in the presence of market
pressures and uncoordinated limit rules among diverse exchanges.

Our paper proceeds as follows. We begin by describing our data in
Section~\ref{data} and provide an analysis of the distributional
characteristics of the data during active and passive subperiods in
Section~\ref{distributions}. In Section~\ref{model}, we describe a
compound multifractal model of asset returns in clock time which
composes a Gaussian model of trade-time returns with a modification of
the MSMD model of \citet{Chen2013}. Section~\ref{results} estimates
the model and compares Monte Carlo simulations with observed
data. Section~\ref{application} extrapolates the model to draw
conclusions regarding market behavior during periods of extremely high
volatility. Section~\ref{conclusion} concludes.

\section{Data}
\label{data}

In this paper we focus our analysis exclusively on the Chicago
Mercantile Exchange (CME) near-month E-mini S\&P 500 Futures contract
(commodity ticker symbol ES). Although the CME provides a variety of
E-mini products, the E-mini S\&P 500 futures contract is the most
heavily traded, and for this reason it is commonly referred to as
\emph{the} E-mini. As indicated by its name, the E-mini is a futures
contract on the value of the S\&P 500 index, with a notional value of
50 times the index. We obtained the full record of tick-by-tick trades
for the period 18 May 2013 to 18 August 2013 by parsing the CME
historical files, encoded in FIX format, which we use to estimate and
evaluate our model in Section~\ref{model} and \ref{results}. We also
obtained similar data for the 27-month period spanned by 27 April 2010
to 17 August 2012, which is used in the out-of-sample analysis
reported in Section~\ref{application}.

Despite the fact that the
E-mini is a futures contract that does not trade on equities
exchanges, its statistical behavior characterizes the dynamics of the
equity markets as a whole. This is attributed to its liquidity and the
relationship of price formation and information transmission between
the futures and equities exchanges in Illinois and New Jersey, as
studied in \citet{Laughlin2014}.

E-mini futures trade Monday through Friday, starting at
5:00 p.m. Central Time on the previous day and ending at 4:15 p.m.,
with an additional daily maintenance trading halt from 3:15 p.m. to
3:30 p.m., Central Time. We aggregate multiple trades occurring within
single milliseconds as unit transactions and assign to them the final,
in-force price of the millisecond as the price of the trade. While not
a perfect approximation, this assumption exploits the fact that
multiple transactions with the same time stamp are often attributable
to a single aggressor order filling several resting orders at the same
price level and also allows us to circumvent singularities associated
with zero durations in our subsequent models. Our resulting data for
the sample period contains a total of 6,832,305 such transactions.

The quoted price of the E-mini corresponds to the index value of the
S\&P 500. Since the S\&P 500 is an index and not a traded asset, its
value is measured in ``points'' rather than dollars, and as a result
E-mini prices are denoted by ``points''. The minimum
tick increment in the quoted price of the E-mini is 0.25 points, which
is the smallest amount by which the E-mini price can change (up or
down). In reality, the minimum E-mini contract size is
\$50 $\times$ the value of the S\&P 500, which means that the actual
traded price is \$50 times the quoted price, with a corresponding
minimum increment of \$12.50. For the remainder of the paper we will
use \emph{quoted} E-mini prices, measured in points, which correspond
directly to the S\&P 500 index value. During May through August of
2013, the S\&P 500 Index traded between $P=1570$ and $P=1710$,
indicating a typical minimum fractional increment in the index price
of $\Delta P/P\sim1.5\times 10^{-4}$.
 
Because we are interested in investigating the distributions of
intra-day asset returns during news event periods and non-event
periods, we consider subsamples of the data that sort according
to news events. Since the E-mini is a futures contract on a market
aggregate, it is almost exclusively affected by major macroecnomic
announcements, and not by smaller scale, industry- or firm-specific
news. For this reason, we classify news events periods with the
EconoDay calendar (\url{econoday.com}), which lists major
pre-scheduled news announcements in the U.S. and which powers
calendars for outlets such as the Wall Street Journal. Since the
majority of pre-scheduled news announcements are made at 8:30 a.m. and 10:00
a.m. we form an event-driven dataset of all E-mini trades that occurred
during a 1000-second (roughly 16-minute) window following a news
announcement at those times. That is, the event-driven dataset is
comprised of trades from approximately 8:30-8:46 a.m. and
10:00-10:16 a.m. that follow any pre-scheduled news
announcement during the sample period. We form a corresponding
non-event-driven subsample that constitutes all trades during the same
time windows on days when announcements were not scheduled. A
1000-second window encompasses participants ranging from the fastest
algorithmic traders to human traders who manually read the news,
consider its implications, and trade on their resulting conclusions.

For the remainder of the paper, we will refer to the event-driven
subsample as the \emph{active} data and the non-event subsample as the
\emph{passive} data. Despite the fact that trading is much heavier
during the time windows of the active data, only about 45\% of the
time windows coincide with scheduled news announcements: there are a
total of 43 news periods and 54 non-news periods over the three-month
sample. As a result, the two datasets are roughly equal in size:
191,127 records in the active data and 174,041 records in the passive
data. Hence, our results below are not an artifact of clock time
(both data sets are drawn from the same intra-day time periods) or
sample size.

\begin{figure}[!htb]
\begin{center}
\includegraphics[width=6in]{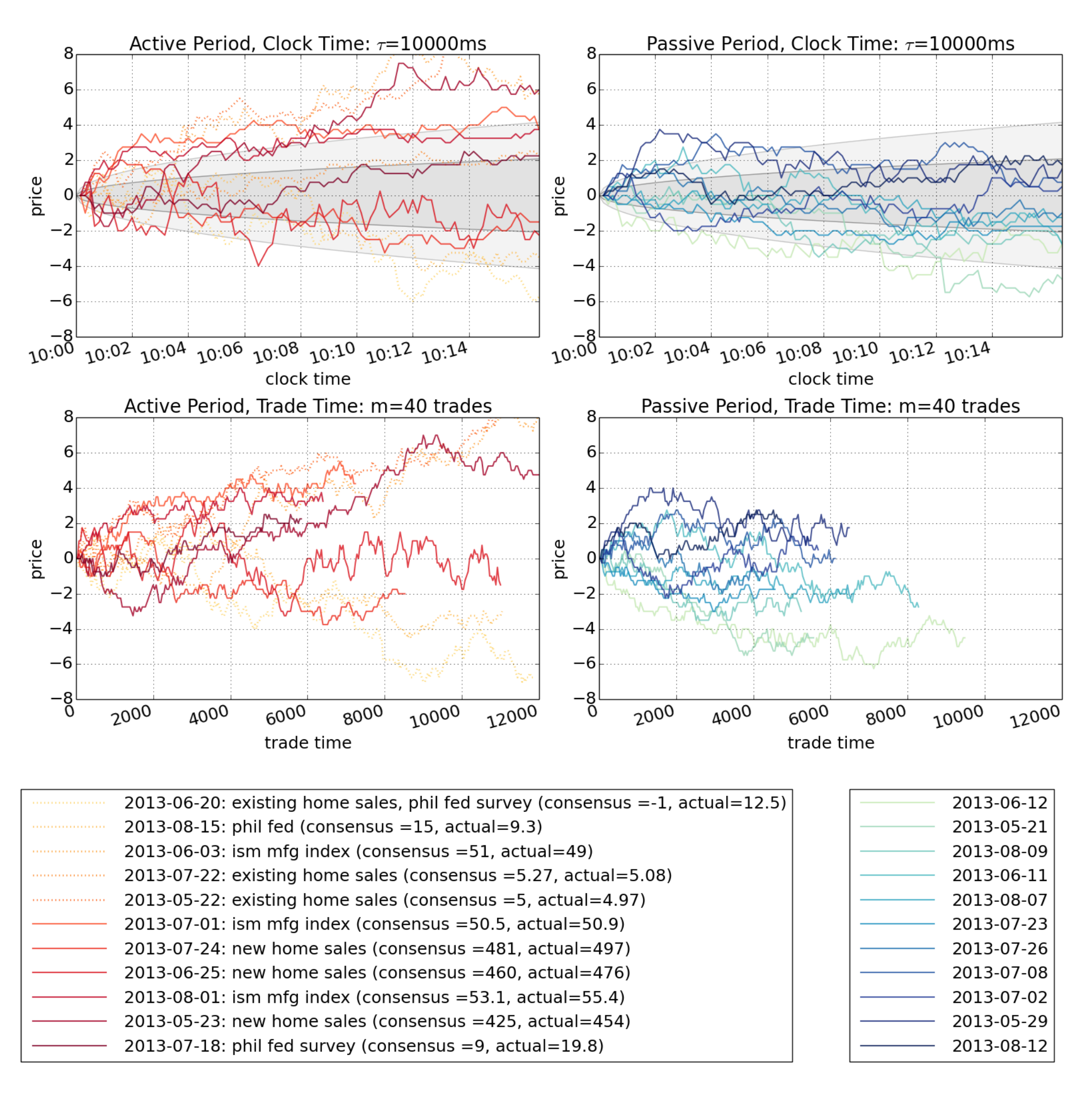}
\caption{Time series plots of E-mini S\&P 500 futures contract prices
  for the 1000 seconds following 10:00 a.m. news announcements (red)
  and the 1000 seconds following 10:00 a.m. for a random selection of
  days without announcements (blue). Prices are all normalized by
  subtracting the initial price from all subsequent values. The upper
  panels show clock-time prices while the lower panels show
  trade-time prices. The legend reports selection dates and
  information on the individual news announcements. The shaded regions
  in the upper panels depict expected $1-\sigma$ Brownian motion
  diffusion bounds on the S\&P 500, obtained by deannualizing Gaussian
  volatilities of $\sigma$=0.1 (inner envelope) and  $\sigma$=0.2
  (outer envelope).}
\label{tsComp}
\end{center}
\end{figure}

Figure~\ref{tsComp} depicts sample paths of E-mini prices during
both active and passive periods of our data set. In
Figure~\ref{tsComp}, and for the remainder of the paper, we use red to
signify active-period data and blue to signify passive-period
data. The upper left panel of the figure shows price series for the
1000 second period following 10:00 a.m. on all days in our sample with
a news announcement at that time. These constitute 11 of the 43 news
event windows; the remaining 32 news announcements occurred at 8:30
a.m. The prices are sampled at 10-second increments and are normalized
by their initial price so that each series begins at the origin. The
shading and texture of the individual series are related to the the
degree by which the actual announcement differed from a prior
consensus forecast. Solid lines indicate announcements that exceeded
the consensus and dotted lines fell short. In addition, we sorted the
data according to percentage deviation from the consensus, using
darker colors to indicate stronger performance and lighter colors to
indicate weaker. The legend briefly describes the date and nature of
each announcement, along with the prior consensus and actual reported
values. As an example, the 10:00 a.m. announcement on 23 May 2013
reported new home sales in the U.S. for the month of April. The prior
forecast was that 425,000 units were sold and the actual reported
value was 454,000 units -- a 6.8\% deviation in excess of the
consensus.

The upper right panel of Figure~\ref{tsComp} shows similar time series
for a random selection of 11 days on which there was no 10:00
a.m. news announcement. The figure legend reports the chosen days. In
this case, shading sorts series according to the magnitude of return
at the end of the 1000 second sample, with darker colors indicating
larger returns and lighter colors indicating smaller returns.

The lower panels of Figure~\ref{tsComp} replicate the time series of
the upper panels, but aggregate prices according to trade time rather
than clock time. As we will formalize in the next section, trade time
measures time increments according to a specific number of trades
occurring rather than a specific amount of wall-clock time elapsing. In
this case, we choose $m=40$, meaning that each unit of time is defined
by 40 trades, and the resulting trade-time series depict the prices
sampled at that frequency.

Two features of the plots are worth highlighting briefly, although we
will address them in more detail in the remainder of the paper. First,
price movements, relative to their starting points, are much more
volatile in the active periods than in the passive periods. The shaded
regions in the upper panels of Figure~\ref{tsComp} depict expected
$1-\sigma$ Brownian motion diffusion bounds on the S\&P 500, obtained
by deannualizing Gaussian volatilities of $\sigma$=0.1 (inner
envelope) and  $\sigma$=0.2 (outer envelope). We note that these
volatilities can be very roughly equated to values for the CBOE
Volatility (VIX) Index of 10 and 20 respectively, which approximately
bounded the U.S. equity market during the 3-month period covered by
our analysis. (The lowest observed closing value for the VIX Index was
11.84 on Aug 5, 2013 and the highest observed closing value was
20.49 on June 20, 2013.) As the figure shows, E-mini prices after news
announcements tend to exceed the expected S\&P 500 volatility, while
those during non-news periods are more likely to fall within the
shaded envelope.

The second feature to note is that the trade-time series are not of
uniform length. This is due to the fact that the number of trades that
occurred in each of the 1000 second time windows was not identical. In
fact, it is apparent from the plots that trade-time series following
reports that beat forecasts are shorter than those which were below
consensus, indicating heavier trading volume on negative news. A
similar feature is observed in the passive-period trade-time panel
(lower right) were trading appears to be heavier during periods in
which prices are declining.

\section{Empirical Distributions of Intra-Day Asset Returns}
\label{distributions}

In this section we emphasize some of the key features of observed
intra-day asset returns distributions. We define clock-time returns as
\begin{equation}
	r_{\tau}(t) = p(t) - p(t - \tau) \label{ctrets}
\end{equation}
where $p(t)$ is the price of an asset at time $t$ and $\tau$ is the
clock-time duration under consideration (such as 1000
milliseconds). An alternative definition of returns can be provided in
trade time, where time increments are measure by a fixed number of
trades:
\begin{equation}
	r_m(n) = p(n) - p(n - m), \label{ttrets}
\end{equation}
where $n$ represents the $n$-th trade and $m$ represents the number of
trades in a unit of time.


Empirical asset returns (measured in clock time) typically exhibit
notable features such as leptokurtosis and conditional
heteroskedasticity, regardless of time scale. Much effort has been
expended over the course of decades to model the heavy tails of
returns distributions (\citet{Mandelbrot1963}) as well as the strong
autocorrelation of volatility (\citet{Engle1982} and
\citet{Bollerslev1986}). In the subsequent analysis we show that the
observations of \citet{Brada1966}, \citet{Mandelbrot1967},
\citet{Clark1973} and \citet{Ane2000}, that trade-time returns are
nearly Gaussian, extend to high-frequency, intra-day returns when
controlling for pre-scheduled news announcements.

\subsection{Intra-day Clock-Time and Trade-Time Distributions}

As reported in Section~\ref{data}, the 43 active 1000-second time
intervals contain 191,127 transactions, while the 54 passive
1000-second time intervals contain a total of 174,041
transactions. This corresponds to an average of 4.44 transactions per
second in the active subsample and 3.22 transactions per second in the
passive subsample. To strike a balance, we assume in the remaining
analysis that 1 tick (transaction) corresponds to 0.25 seconds, or 250
milliseconds.  We consider time aggregated returns for $\tau = \{250,
500, 1000, 4000, 10000, 30000\}$ milliseconds (i.e. our largest time
scale is 30 seconds) and $m = \{1,2,4,40,400,4000\}$ trades, which
according to our approximation are roughly corresponding time
intervals.

\begin{figure}[ht]
\begin{center}
\begin{mdframed}
\includegraphics[width=6in]{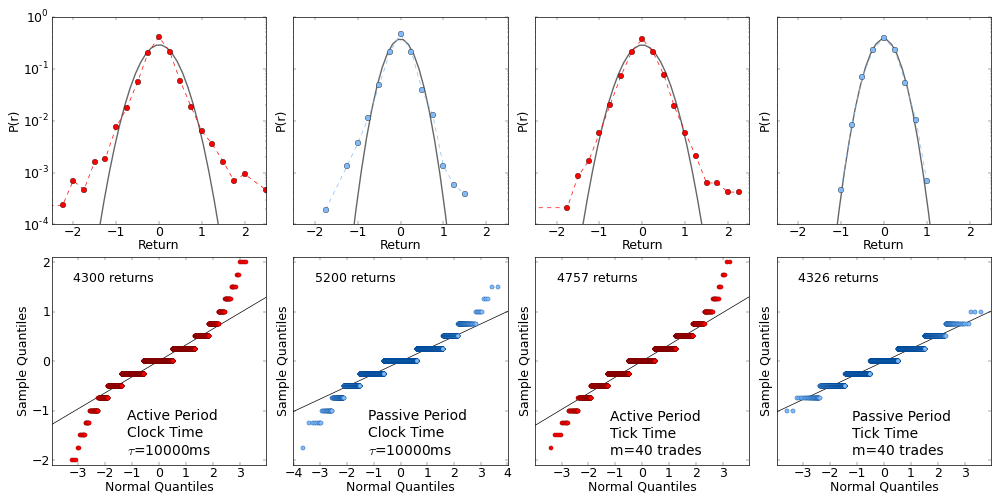}
\end{mdframed}
\caption{Empirical density and Q-Q plots for clock-time and trade-time
  returns for both active and passive subsamples. The clock-time
  interval is $\tau = 10000$ milliseconds (10 seconds) and the
  trade-time interval is $m = 40$ (chosen to roughly match the
  clock-time interval).}
\label{qqHistComp}
\end{center}
\end{figure}

Figure~\ref{qqHistComp} isolates empirical returns distributions for a
single time scale during both active and passive news regimes: $\tau =
10000$ ms (10 seconds) and $m = 40$ trades. The upper row of plots
depict the empirical density functions of the discretely observed
returns in our samples, superimposed upon Gaussian distributions that
are estimated by maximum likelihood. These are shown with the vertical
axis on a log scale in order to highlight discrepancies in the tails
of the distributions.
The second row of plots in Figure~\ref{qqHistComp} show corresponding
quantile-quantile (Q-Q) plots of the distributions in the upper
panels. Q-Q plots compare the quantiles of two distributions and are
an powerful way to visualize points of departure between them without
transforming the data to a logarithmic scale as we did in the upper
panels. Hence, the bottom row plots the empirical quantiles of each
dataset against the theoretical quantiles of the Gaussian distribution
obtained via maximum likelihood (although the quantiles on the x-axis
have been standardized). If the theoretical and empirical densities
coincide, their quantiles will exhibit a linear
relationship. Departures are highlighted via nonlinearities in the Q-Q
plot.

As we will see in more detail below, Figure~\ref{qqHistComp} shows
that clock-time intra-day returns have heavy tails for both active
and passive time periods, which indicates that the probability
distribution generating the data has higher probabilities of extreme
events than a Gaussian distribution (even during passive, non-event
periods). The same is true of active-period trade-time returns. The
characteristic heavy tails of the empirical distributions are visible
in the Q-Q plots as a convex-concave departure from
linearity. Remarkably, the empirical density and corresponding Q-Q
plot in the last column of Figure~\ref{qqHistComp} show that
passive-period, trade-time returns conform very closely to a Gaussian
distribution -- they do not have the same propensity for extreme
events as the other datasets. This is a key result of our analysis,
and we analyze it in more detail in the forthcoming sections. For the
remainder of this section, we will depict distributional differences
via Q-Q plots.

\begin{figure}[ht]
\begin{center}
\includegraphics[width=6.5in]{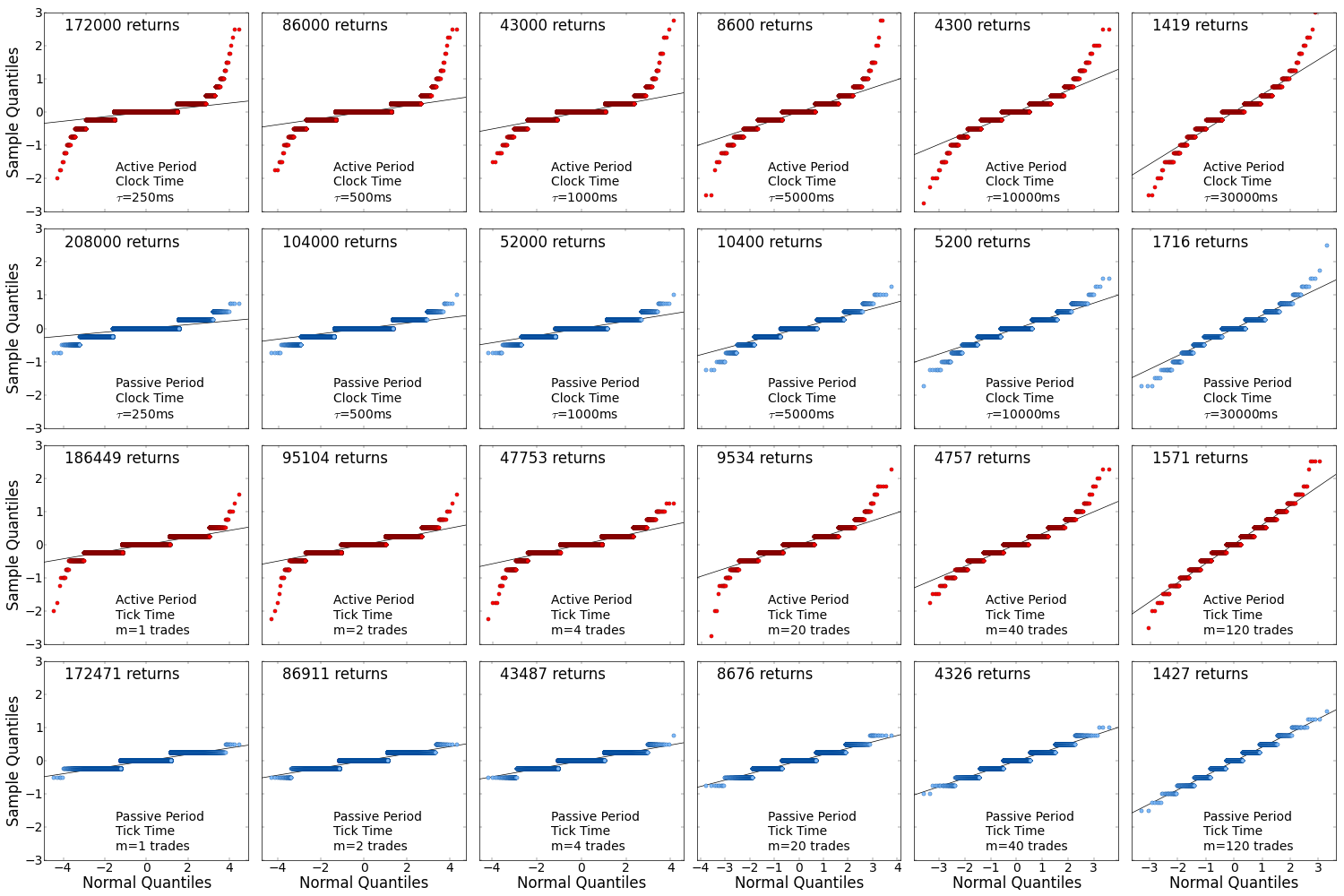}
\caption{Sample Q-Q plots for returns in both
  clock time and tick time and in both active and passive
  subsamples. Each panel corresponds to a clock-time or trade-time
  interval during either the active or passive subsample of data. The
  vertical axes depict sample quantiles of the data and the horizontal
  axes depict theoretical quantiles of best-fit Gaussian
  distributions.}
\label{qqAll}
\end{center}
\end{figure}

Figure~\ref{qqAll} shows Q-Q plots of clock-time and trade-time
returns during active and passive periods for all time scales that we
consider.  The panels in the first and second rows of the figure
compare the empirical quantiles of the clock-time returns with the
theoretical quantiles of the best-fit Gaussian distributions for
active and passive periods, respectively. The panels in the third and
fourth rows are the same for trade-time returns.

Not surprisingly, the upper rows of Figure~\ref{qqAll}
show that clock-time returns distributions are markedly different from
a Gaussian density, over a variety of intra-day time scales. In fact,
the patterns exhibited in each of the panels are indicative of heavy
tails, especially at fine time scales, while as $\tau$ increases the
leptokurtosis diminishes. In addition to the diminishing tail weight
with increasing $\tau$, comparison of active-period and passive-period
clock-time returns shows that active, news-event periods have much
heavier tails. This is exactly as we would anticipate, since periods
following news announcements should have a higher frequency of large
price movements.

The third row of Figure~\ref{qqAll} shows that trade-time returns
during the active subsample are quite similar to those of clock-time
returns during active and passive subsamples: leptokurtosis
diminishing with $m$. The final row of the figure, however, highlights
the surprising result noted above: during passive, non-event time
periods, trade-time returns conform quite closely to a Gaussian
distribution. We emphasize that this empirical result has very
important asset pricing implications. In fact, Gaussianity
of trade-time returns, controlling for known news announcements, will
be foundational for the model and results we present in
Sections~\ref{model} and \ref{results}.

Gaussianity of trade-time returns is not immediately apparent
when studying intra-day data because trading during
news event periods has a large impact on the unconditional
distribution of returns. For considerations of space, we have not
included Q-Q plots for the full sample of data (without sorting on
news events), however the corresponding empirical distributions look
quite similar to those we have shown for the active subsample. That
is, trading during limited periods of pre-scheduled news announcements
has a very large impact on the tail probabilities of the unconditional
returns distributions. In contrast, as the bottom panels of
Figure~\ref{qqAll} indicate, if one only considers returns during
non-event times, a Gaussian distribution provides an excellent
fit. Knowing that asset returns accumulated in trade time follow a
Gaussian distribution during quiescent periods of market activity is
a very powerful result.

\begin{figure}[ht]
\begin{center}
\includegraphics[width=6.5in]{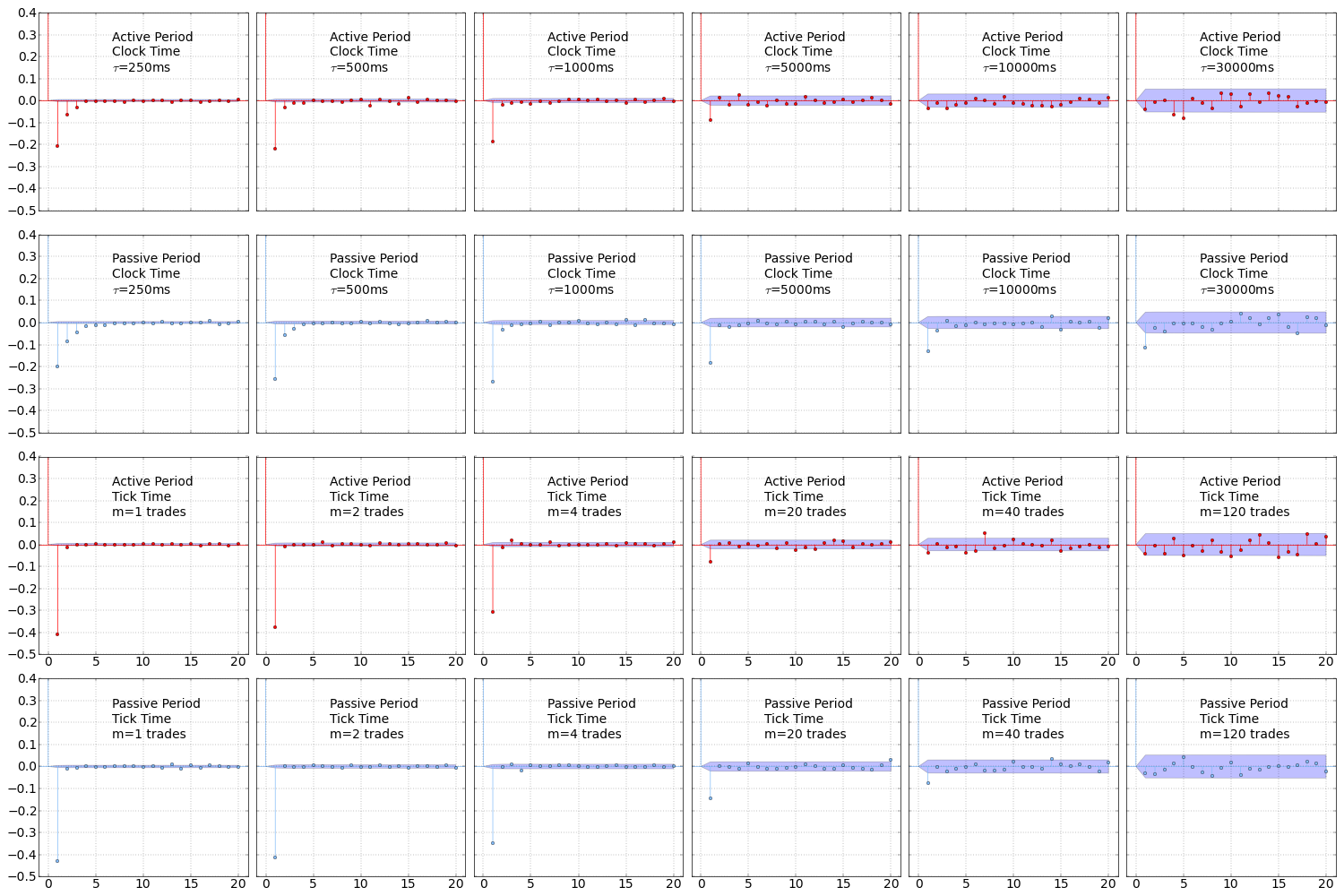}
\caption{Sample autocorrelation plots for returns in both
  clock time and trade time and in both active and passive
  subsamples. Each panel corresponds to a clock-time or trade-time
  interval during either the active or passive subsample of data.}
\label{acAll}
\end{center}
\end{figure}

\begin{figure}[ht]
\begin{center}
\includegraphics[width=6.5in]{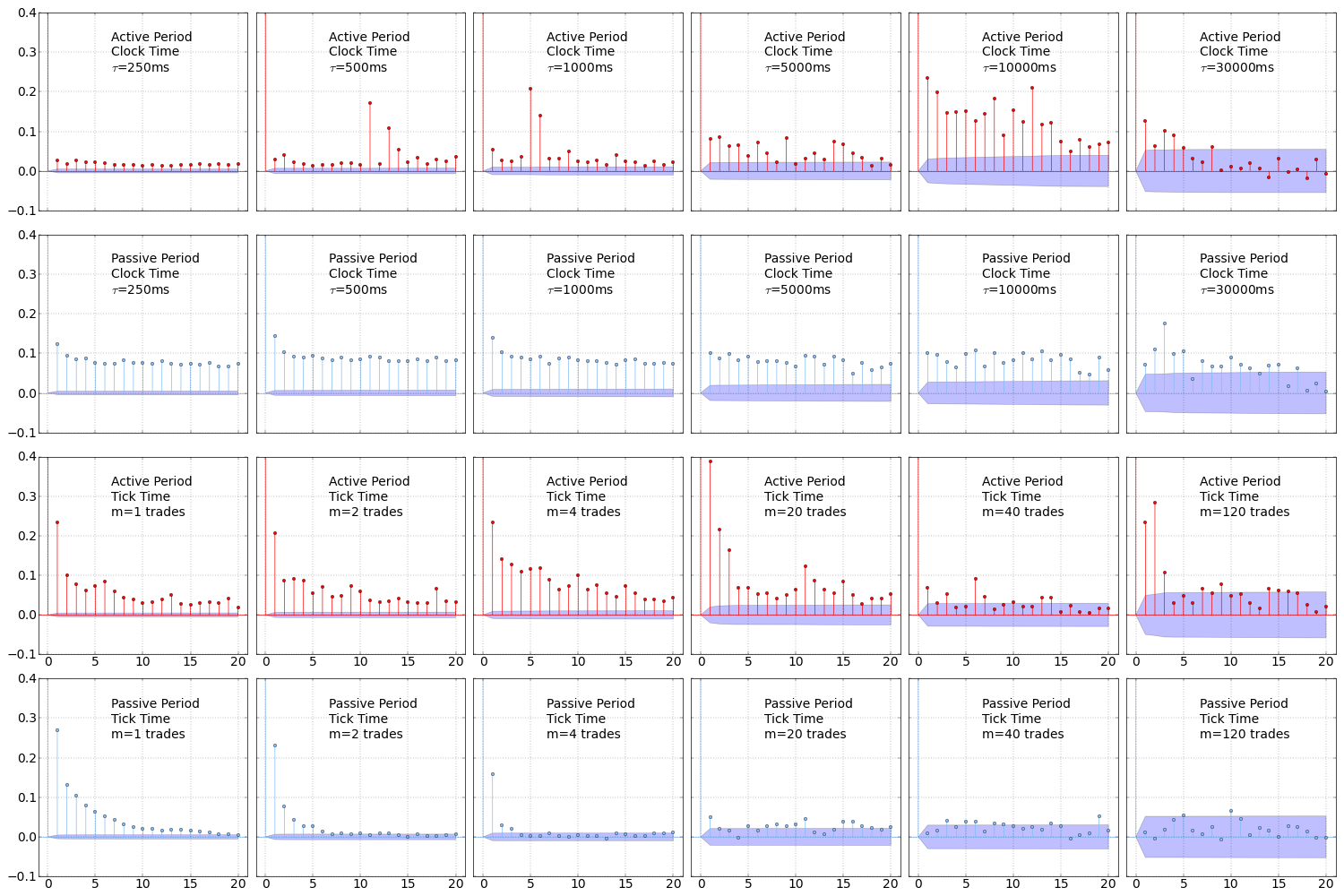}
\caption{Sample autocorrelation plots for squared returns in both
  clock time and trade time and in both active and passive
  subsamples. Each panel corresponds to a clock-time or trade-time
  interval during either the active or passive subsample of data.}
\label{ac2All}
\end{center}
\end{figure}

Figures~\ref{acAll} and \ref{ac2All} depict sample autocorrelation
functions (ACFs) of returns and squared returns (respectively) for
the clock-time and trade-time intervals considered in the Q-Q plots of
Figure~\ref{qqAll}. It is generally accepted that asset returns
exhibit no significant autocorrelation except at very fine time
scales, where bid/offer bounce and mean reversion induce negative
correlation among adjacent trades. Figure~\ref{acAll} corroborates
these stylized facts: at all but the finest time scales and lowest
lags, returns show no autocorrelation. For both active and passive
periods, autocorrelations of trade-time returns appear to be more
muted than those of clock-time returns. In fact, for $\tau=250$ms and
$\tau=500$ms, clock-time returns exhibit significant and decaying
autocorrelation beyond the first lag, a feature which is not present
in the trade-time data, or at least only mildly so for $m=1$ (where
there appears to be a very small, significant lag-2 autocorrelation).

ACFs of squared returns, shown in Figure~\ref{ac2All}, are typically
used to diagnose the persistence of volatility, which is a well
documented property of financial asset returns. In particular, 
the panels in the second row of Figure~\ref{ac2All}, which correspond
to passive-period, clock-time returns, are exemplary of the persistent
nature of volatility. The remaining panels show that active-period
returns (both in clock time and in trade time) exhibit volatility
peristence, but of a somewhat less consistent nature, and that
passive-period, trade-time returns only exhibit a small degree
volatility persistence at fine time scales, with the persistence
quickly diminishing as $m$ increases. This latter results is a feature
of the data will allow us to approximate passive-period, trade-time
returns as independent draws from a Gaussian distribution and will
facilitate the task of building a Monte Carlo approximation of the
clock-time returns distribution in Section~\ref{results}.

\section{Model}
\label{model}

In this section we develop a hierarchical model of clock-time
returns that mixes a distribution of trade-time returns with a
distribution of trade arrivals. For a given trade-time interval $m$,
we assume that trade-time returns are independently drawn from
a Gaussian distribution,
\begin{align}
r_m(n) \stackrel{i.i.d.}{\sim} \mathcal{N}(\mu, \sigma), \forall
n. \label{gaussTick}
\end{align}
According to Equation~\eqref{gaussTick}, when $m=1$ the price
differences between \emph{each} transaction result in Gaussian
returns. However, our notation generalizes to allow price differences
across every $m$ transactions to follow a Gaussian distribution, which
is direct result of Gaussian aggregation.

Trade arrivals will be distributed according to a counting process to
be specified later. For a given clock-time duration $\tau$, we will
denote the number of $m$-period executed trades as $N_m(\tau)$, with
corresponding probability $P(N_m(\tau) = k)$, for $k =
\{0,1,2,\ldots\}$. If we only observe returns that are aggregated over
time interval $\tau$, then we are interested in the distribution of
the random variable
\begin{align}
  r_{\tau}(t) = \sum_{i=1}^{N_m(\tau)} r_m(n).
\end{align}
The probability density function of $r_{\tau}(t)$ is,
\begin{align}
  p(r_{\tau}(t)|\mu,\sigma) & = \sum_{k=1}^{\infty} p\left(\sum_{i=1}^k
  r_m(n) \bigg|N_m(\tau)=k, \mu, \sigma \right) P(N_m(\tau) = k). \label{mix}
\end{align}
This distribution is characterized as a finite Gaussian mixture model
with mixture weights that vary according to the probability
distribution of $N_m(\tau)$. The model can also be characterized as a
two-stage hierarchical model in which a number of trades is drawn from
the distribution of $N_m(\tau)$ in the first stage and a single
$\tau$-period return is drawn from the Gaussian distribution of
$r_{\tau}(t) = \sum_{i=1}^{N_m(\tau)} r_m(n) \sim
\mathcal{N}\left(N_m(\tau)\mu, \sqrt{N_m(\tau)} \sigma \right)$ in the
second stage. In the remainder of this section, we consider special
cases of Equation~\eqref{mix} under differing assumptions for the
distribution of $N_m(\tau)$.

\subsection{Compound Poisson Process}
\label{compPois}

A starting point for modeling trade arrivals would be to assume they
follow a Poisson process:
\begin{gather}
  N_m(\tau) \sim \rm{Poisson}(\gamma \tau) \nonumber
  \intertext{or}
  P\left(N_m(\tau) = k\right) = \frac{(\gamma \tau)^k}{k!}
  \exp\{-\gamma \tau\},
\end{gather}
where $\gamma$ is the arrival intensity parameter. In this case
Equation~\eqref{mix} becomes
\begin{align}
  p(r_{\tau}(t)|\mu,\sigma) & = \sum_{k=1}^{\infty} \frac{1}{\sigma
    \sqrt{2 \pi k}} \exp\left\{-\frac{1}{2} \frac{(\sum_{i=1}^k r_m(n)
    - k\mu)^2}{k \sigma^2}\right\} \times \exp\{-\gamma \tau\}
  \frac{(\gamma \tau)^k}{k!}. \label{mixPoisson}
\end{align}
This is a version of the compound Poisson process developed by
\citet{Press1967} and \citet{Press1968}. Unfortunately, its density
function cannot be obtained in
closed form. However, the density can be easily approximated by Monte
Carlo simulation: first making independent draws from the Gaussian
distribution and then accumulating random numbers of those Gaussians
according to integer deviates drawn from the Poisson
density. Alternatively, in some cases we can approximate
Equation~\eqref{mixPoisson} with an analytical formula: when $\tau$ is
large relative to $\gamma$, the $\mathcal{N}(\gamma \tau, \sqrt{\gamma
  \tau})$ density serves as  good approximation to the
$\rm{Poisson}(\gamma \tau)$ density. In this special case,
$r_{\tau}(t)$ is also approximately distributed as a Gaussian
(compounding two Gaussian densities together in this manner results in
another Gaussian density). However, for small values of $\tau$ (when
the analytic Gaussian approximation does not hold), the resulting
distribution of $r_{\tau}(t)$ exhibits leptokurtosis. Thus, as we will
see below, Poisson trade arrivals over short time horizons are able to
partially explain extreme events for asset returns although they are
inadequate for explaining volatility clustering.

It is important to note that the assumption of Poisson trade arrivals
implies that inter-trade durations are distributed as an Exponential
random variable, with rate $\gamma$. In Section~\ref{results} we will
estimate the parameters of this model. We will see that although the
Exponential model of trade durations (Poisson model of trade arrival)
can partially explain the heavy tails of returns distributions 
and volatility clustering, it is under-dispersed relative to the
actual distribution of trade durations and cannot fully account for
those features of the data. Hence, a better model of inter-trade
duration is needed to adequately explain observed data.

\subsection{Compound Multifractal Process}
\label{compMF}

\citet{Chen2013} develops a model for inter-trade durations that more
adequately explains the dynamics of intra-day trade arrivals. The
model of that paper, the Markov-switching multifractal duration (MSMD)
model, builds on the multifractal volatility model of
\citet{Calvet2001}, \citet{Calvet2002} and \citet{Calvet2004}. In
particular, the authors adapt the multifractal volatility model to
explain trade durations (rather than volatility) as aggregates of
latent shocks. They demonstrate that the MSMD does quite well at
explaining intra-day trade data for a variety of equities traded in
1993.

The core components of the MSMD model are a set of $\bar{k}$ latent
state variables, $M_{k,i}$, that obey a two-state Markov-switching
process with varying degrees of persistence, $\gamma_k$, for $k = 1,
2, \ldots, \bar{k}$. That is, the distribution of trade durations,
$d_i$, is governed by the equations,
\begin{subequations} \label{msmd}
\begin{gather}\
  d_i = \frac{\eps_i}{\lambda_i} \label{msmd1} \\
  \eps_i \sim Exp(1) \label{msmd2} \\
  \lambda_i = \lambda \prod_{k=1}^{\bar{k}} M_{k,i} \label{msmd3} \\
  M_{k,i} = \begin{cases} M & \text{with probability} \,\, \gamma_k
    \\ M_{k,i-1} & \text{otherwise} \end{cases} \label{msmd4} \\
  \gamma_k = 1 - (1-\gamma_{\bar{k}})^{b^{k-\bar{k}}} \label{msmd5} \\
  M = \begin{cases} m_0 & \text{with probability} \,\, 1/2 \\ 2-m_0 &
    \text{otherwise}. \end{cases} \label{msmd6} 
\end{gather}
\end{subequations}
Hence, the MSMD can be succinctly characterized by five parameters:
$\bar{k} \in \mathbb{N}$, $\lambda > 0$, $\gamma_{\bar{k}} \in
(0,1)$, $b \in (1,\infty)$ and $m_0 \in (0,2]$. The intuition is that
  conditional on knowing the values of the latent state variables,
  inter-trade durations are Exponentially distributed with intensity
  parameter $\lambda_i$. However, as time evolves, the latent states,
  $M_{k,i}$, switch values with varying degrees of persistence,
  $\gamma_k$. This causes the unconditional distribution of trade
  durations to be a mixture of Exponentials, which is consistent with
  the over-dispersion property of observed data, described in
  Section~\ref{comparison}. The latent states can be interpreted as
  shocks that have varying impacts over diverse timescales, some
  having short-horizon and others have long-horizon effects. The value
  $b$ governs a tight relationship between the persistence parameters,
  $\gamma_k$, and is responsible for the parsimony of the model: even
  with a large number of latent states, $\bar{k}$, the model is always
  characterized by a total of five parameters. The choice of $b$
  dictates the degree of heterogeneity in values of persistence
  parameters. For more insight regarding the MSMD model, see
  \citet{Chen2013} and \citet{Calvet2008}.

\begin{figure}[ht]
\begin{center}
\includegraphics[width=5.8in]{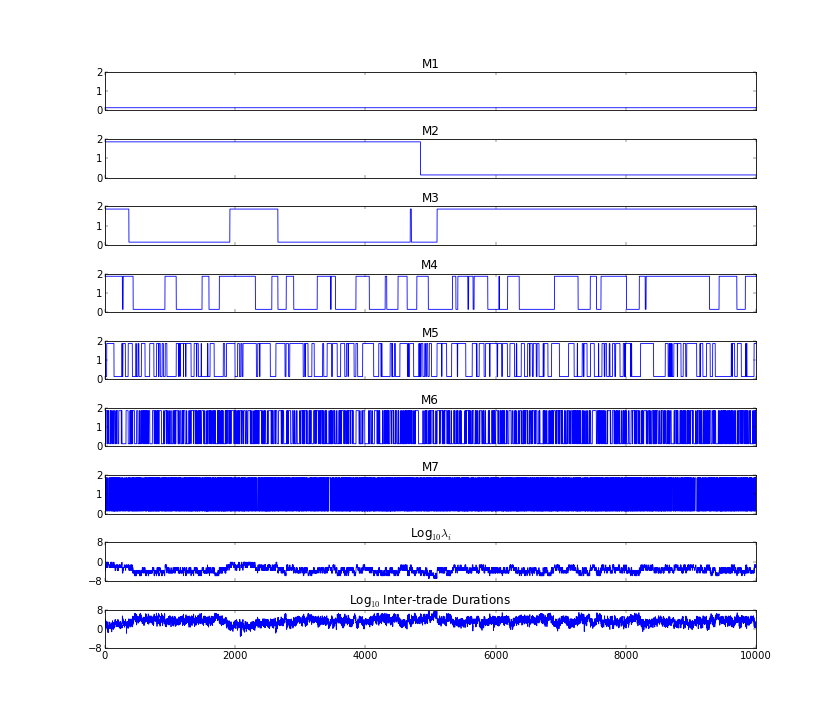}
\caption{A simulation of MSMD inter-trade durations for $\bar{k} =
  7$. The upper panels depict the time paths of latent states and MSMD
intensity parameter and the bottom panel depicts the durations
themselves. The layout of this figure is the same as that presented in
Figure 5 of \citet{Chen2013}.}
\label{chenFig5}
\end{center}
\end{figure}

Figure~\ref{chenFig5} is analogous to Figure~5 in
\citet{Chen2013}. It shows a simulation of MSMD durations (using the
$\bar{k} = 7$ parameter values in Table~\ref{msmdEstimates}) and the
associated time paths of the latent states as well as the composite
intensity parameter $\lambda_i$. The penultimate panel of the figure
depicts the stochastic intensity parameter (on a log scale), which is
driven by the latent states that switch at differing frequencies. This
contrasts with a simple Exponential model of inter-trade durations
which fixes a constant intensity parameter. The result is that MSMD
durations exhibit far greater heterogeneity than those of the
constant-intensity Exponential. \citet{Chen2013} liken stochastic
intensity in duration models to stochastic volatility in returns
models: ``Just as stochastic volatility `fattens' Gaussian conditional
returns distributions, so too does MSMD `over-disperse' exponential
conditional duration distributions.'' (p. 9). In fact, as noted below,
we find that stochastic intensity plays the dual role of
duration over-dispersion and returns tail fattening.

Just as the Exponential model of \emph{inter-trade durations} is tied
to the Poisson model of \emph{trade arrival}, the MSMD model is also
tied to a model of trade arrival. As with the Poisson, specifying the
probabilities $P(N_m(\tau)=k)$, for $k = 0,1,2,\ldots$, corresponding
to the MSMD model would also result in a distribution for clock-time
returns that cannot be obtained in closed form. However, unlike the
compound Poisson process, we cannot obtain a closed-form solution for
the counting distribution itself which serves as mixture weights in
the Gaussian mixture model. Fortunately, we can approximate the
counting distribution via Monte Carlo simulation. We show an example
of such an approximation in Section~\ref{comparison}.

The distribution of $r_{\tau}(t)$ retains its hierarchical structure
under the MSMD model, with the number of trades per unit of time being
drawn from a mixture of Poisson distributions in the first stage. We
refer to $r_{\tau}(t)$ as a \emph{compound multifractal process} when
the Gaussian mixture weights correspond to count probabilities
associated with MSMD durations. The strength of the compound
multifractal process lies in the variability of the underlying
counting process: the dispersion of probability over a greater variety
of counts, relative to the simple Poisson model, induces greater
heterogeneity in the Gaussian mixture, which generates fatter tails
for $r_{\tau}(t)$. Intuitively, the random variable $r_{\tau}(t)$
switches between a greater variety of differing sums of Gaussians with
higher probability. In addition, the compound multifractal model
explicitly generates volatility persistence by producing
autocorrelation in the inter-trade duration distribution, via
Markov-switching latent states.

\subsection{Compound Truncated Multifractal Process}
\label{compTMF}

We augment the duration models considered above with a modification of
the MSMD model. Specifically, we truncate the MSMD model by mixing it with
an Exponential model that is fit so that its expected maximal value
is equal to the maximal duration observed in the data. We will provide
more detail on the fitting procedure in
Section~\ref{results}. Mathematically,
\begin{subequations} \label{msmdExp}
\begin{gather}
  d_i = \min\{d_{MSMD}, d_{\overline{Exp}}\} \label{msmdExp1} \\
  d_{MSMD} \sim MSMD(\bar{k}, \lambda, \gamma_{\bar{k}}, b,
  m_0) \label{msmdExp2} \\
  d_{\overline{Exp}} \sim Exp(\nu_{max}), \label{msmdExp3}
\end{gather}
\end{subequations}
where Equation~\eqref{msmdExp2} denotes that $d_{MSMD}$ follows an
MSMD process outlined in System~\eqref{msmd} and $\nu_{max}$ is chosen
so that $\E\left[\max\{d_{\overline{Exp}}\}\right]$ equals the maximum
duration observed in the data.

The truncated process can be conceptualized as combining
two types of traders that are characterized by differing statistical
processes. The orders of one type follow the counting process
associated with the MSMD model and comprise the majority of trading
on exchanges. The orders of the second type follow a simple Poisson
event clock and only result in trades when there is a sufficiently
long duration between orders of the first type. Intuitively, these
types can be thought of as algorithmic traders (following the MSMD
counting process), which compete at low latency for order
execution, and human traders (following the Exponential), which act
independently and exogenously relative to the bulk of trading, and
whose orders arrive much less frequently (at longer durations).

As with the MSMD model, the truncated MSMD model is associated with a
counting process that can only be obtained via Monte Carlo
approximation. Likewise, using the associated truncated MSMD counting
process in Equation~\eqref{mix} gives rise to process for
$r_{\tau}(t)$ that we refer to as the \emph{compound truncated
  multifractal process}. We will show in Sections~\ref{comparison} and
\ref{results} that this latter model provides a much better
description of observed high-frequency returns data.

\subsection{Comparison of Duration Models}
\label{comparison}

We now give a brief comparison of the estimated duration models,
before describing the estimation results in detail in
Section~\ref{results}. This serves to motivate later discussion and
to highlight the strengths and weaknesses of the model components.

\begin{figure}[ht]
\begin{center}
\includegraphics[width=5.25in]{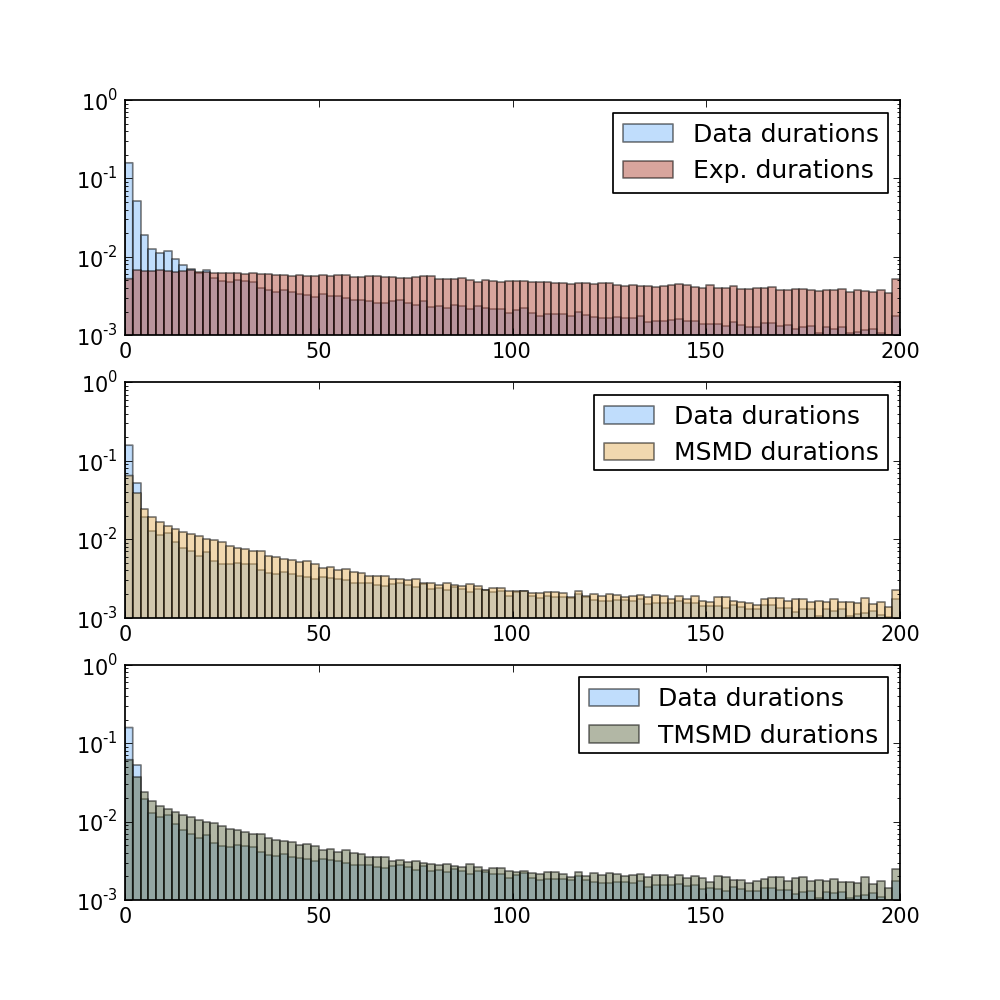}
\caption{Histograms of simulated durations under the Exponential, MSMD
  and TMSMD models as well as observed inter-trade durations in the
  passive-period E-mini returns data.}
\label{durHist}
\end{center}
\end{figure}

Figure~\ref{durHist} shows histograms of 500,000 simulated durations
from each of the three models, using the estimated parameters reported
in Table~\ref{estimates} (described in Section~\ref{results}). Each
panel depicts the histogram of simulated durations from one model
overlaid with the histogram of durations observed in the
passive-period E-mini data. We truncate the histograms at 200 ms to
focus attention on the regions of greatest probability mass. In
addition, to highlight discrepancies in the tails of the
distributions, we have shown the vertical axis on a $\log_{10}$
scale.
It is apparent from the figure that the Exponential duration model
provides a very poor fit to the distribution of the data relative to
the MSMD and truncated MSMD (TMSMD) models. In fact, one important
feature exhibited by the latter two models, as in the data, is
over-dispersion: that variance of inter-trade durations exceeds their
expected value. Under the Exponential model, $\E[d_i] =
\Var(d_i)$. However, the asymmetry and long right tail of the
distribution of the data yield the empirical result that $\E[d_i] <
\Var(d_i)$. This feature is discussed in detail by
\citet{Chen2013}.

\begin{figure}[ht]
\begin{center}
\includegraphics[width=6in]{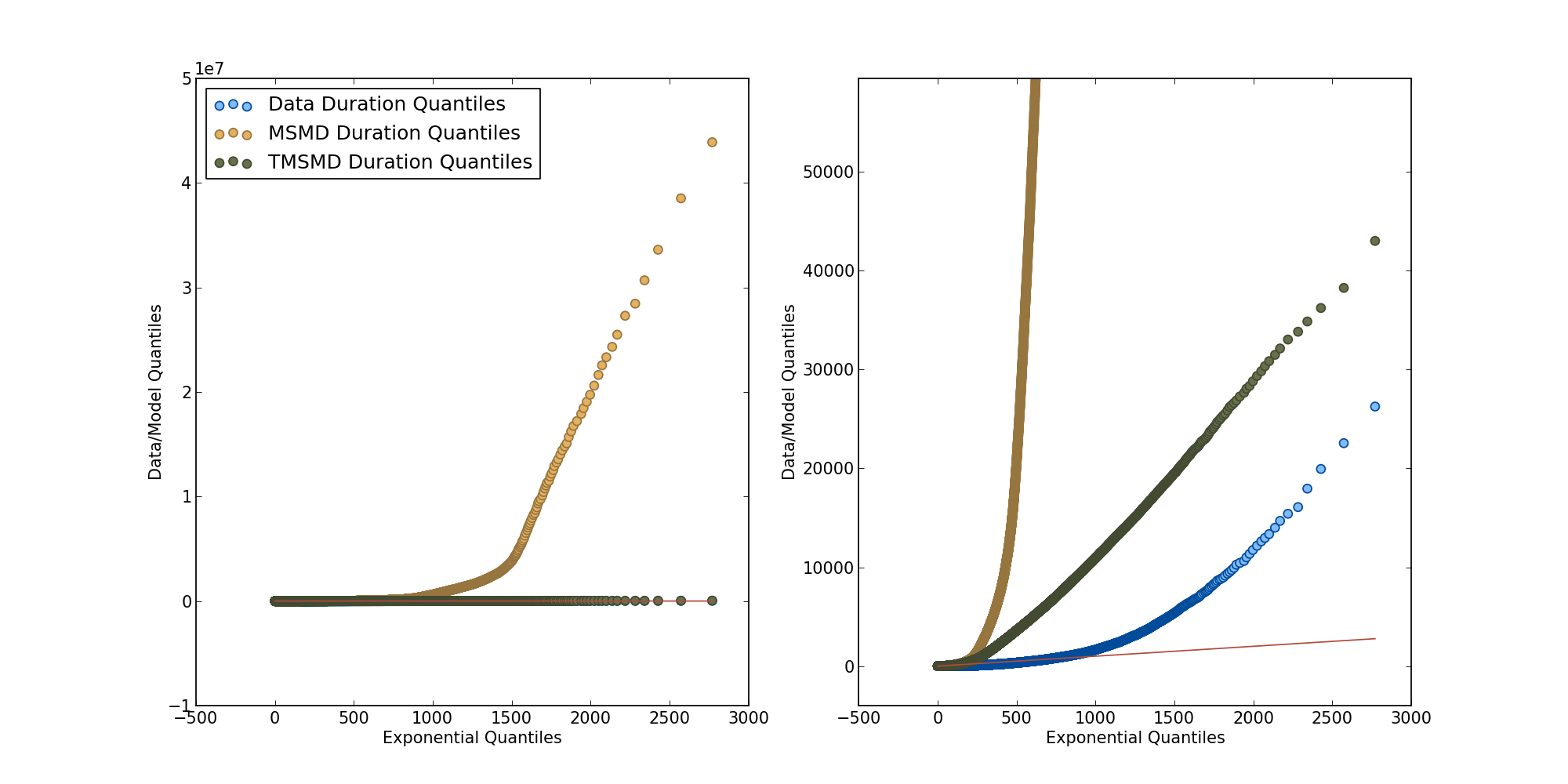}
\caption{Quantile-Quantile plot. The vertical axis shows quantiles of
  simulated durations under the MSMD and TMSMD models and observed
  inter-trade durations in the passive-period E-mini returns data. The
  horizontal axis shows quantiles of simulated durations under the
  Exponential model. All model parameters are taken from
  Table~\ref{estimates}.}
\label{durQQ}
\end{center}
\end{figure}

Figure~\ref{durQQ} shows Q-Q plots of passive-period observed
durations and simulated MSMD and TMSMD durations against simulated
Exponential durations. The second panel is a replicate of a the first
but with a reduced scale on the vertical axis. Most important to note
from the figure is that the empirical distribution of the data has a
much heavier tail than the Exponential distribution. This feature is
responsible for the over-dispersion discussed above. Further, both the
MSMD and TMSMD models produce duration distributions with heavy tails,
but the leptokurtosis of the MSMD distribution is excessive relative
to that observed in the data. The TMSMD distribution, on the other
hand, provides a much closer fit.

\begin{figure}[ht]
\begin{center}
\includegraphics[width=6in]{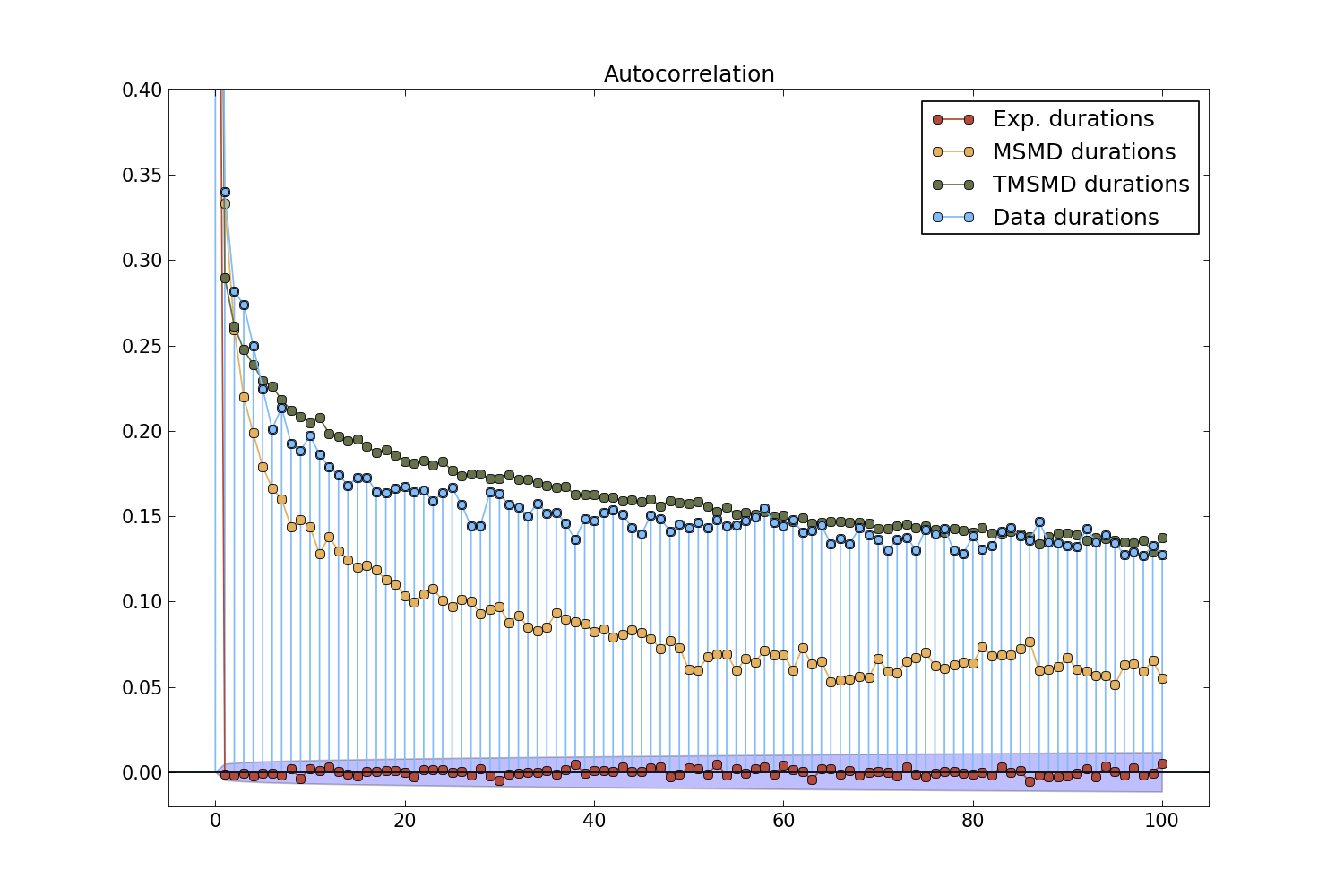}
\caption{Sample autocorrelation functions of simulated durations under
  the Exponential, MSMD and TMSMD models as well as observed
  inter-trade durations in the passive-period E-mini returns data.}
\label{durACF}
\end{center}
\end{figure}

Another important feature of an inter-trade duration model is the
structure of autocorrelations. Figure~\ref{durACF} shows sample
autocorrelations computed for 100 lags for each of the model
simulations as well as the passive-period E-mini data. As noted by
\citet{Chen2013}, observed inter-trade durations exhibit slow decay in
their autocorrelation structure, suggesting long-memory properties. In
contrast, the Exponential model exhibits no autocorrelation since
durations are drawn in an i.i.d. fashion from that model. On the other
hand, the MSMD and TMSMD models provide a much better fit to the
sample autocorrelation function of the data, with the TMSMD strikingly
close. Matching both the density function and the autocorrelation
function of durations is crucial to generating the dynamics observed
in clock-time returns.

\begin{figure}[ht]
\begin{center}
\includegraphics[width=4.75in]{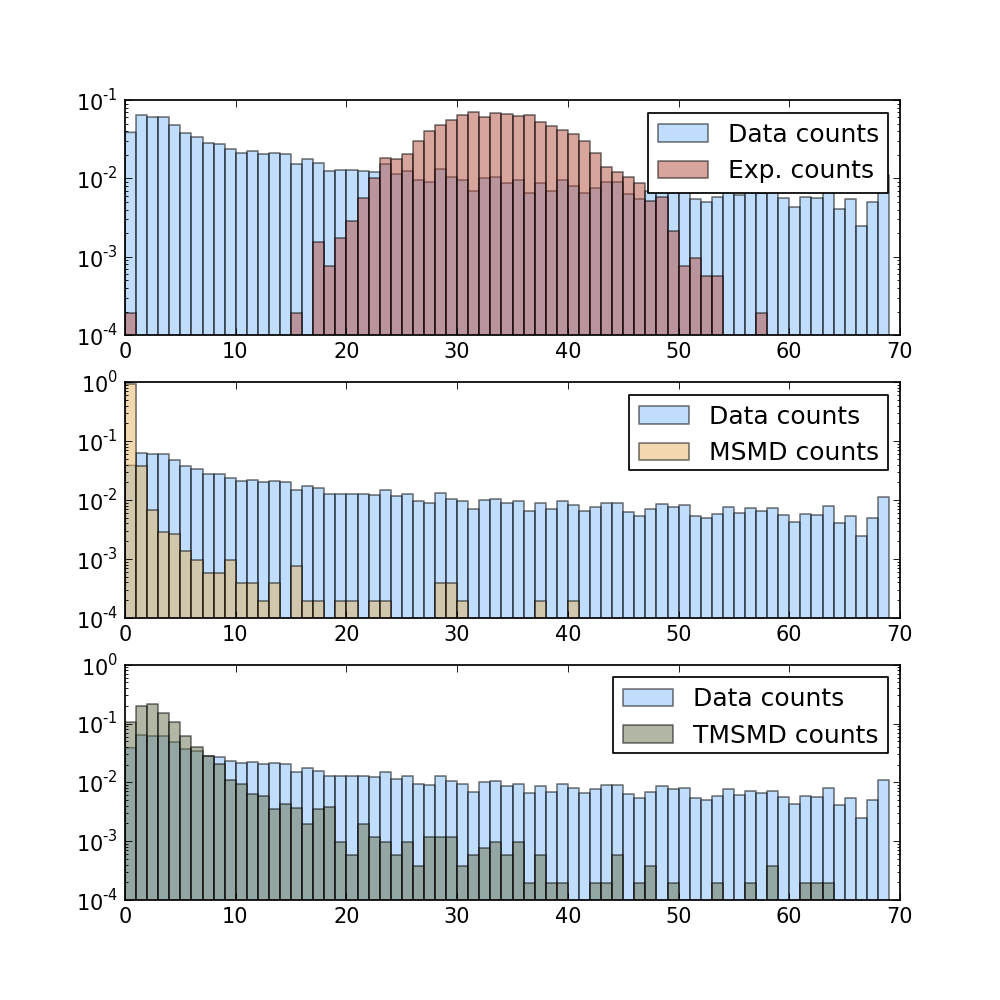}
\caption{Counting process distributions for $\tau = 10000$ ms. Each
  panel shows the counting distribution associated with one of the
  simulations (Exponential, MSMD and TMSMD), described in this
  section, overlaid with the empirical counting density of the
  passive-period E-mini returns. The simulations were obtained using
  the parameters estimates reported in Table~\ref{estimates}.}
\label{countingDens}
\end{center}
\end{figure}

Finally, Figure~\ref{countingDens} shows the counting densities
associated with the duration densities of Figure~\ref{durHist} and a
clock-time interval $\tau = 10000$ ms. That is, using the data and
simulations from each of the models, we obtained respective frequency
counts of trade arrivals within consecutive 10000 ms time
intervals. To highlight tail discrepancies, we have again shown the
vertical axis on a $\log_{10}$ scale. These are approximations of the
actual densities used in Equation~\eqref{mix} to obtain the various
compound distributions of clock-time returns in
Sections~\ref{compPois}-\ref{compTMF}. Each panel overlays the density
associated with a duration model with that of the data. As we
mentioned in Section~\ref{compPois}, the counting process associated
with the Exponential duration model is a Poisson process, which is
depicted in the first panel of Figure~\ref{countingDens}. Clearly, the
Poisson distribution provides a very poor fit to the data, especially
in regions associated with low and high counts. In this particular
case the value of $\tau$ is large enough relative to $\lambda$ that
the density conforms closely to a Gaussian distribution (recall the
the vertical axis is in $\log_{10}$ units). As a result, we expect
that the compound Poisson process, which is a mixture of Gaussian and
Poisson densities, will not fit the data well and that in this special
case they will conform closely to a Gaussian distribution. In
contrast, the counting processes associated with MSMD and TMSMD
durations provide a much better fit, although their right tails
attenuate too quickly relative to the data. Once again, the TMSMD
model provides a much better fit to the empirical density of the
data.

\section{Estimation and Results}
\label{results}

Using the quiescent-market E-mini returns data described in
Sections~\ref{data} and \ref{distributions}, we estimate the
parameters of the component distributions in Equation~\eqref{mix} and
simulate from the mixture model. In particular, we first estimate the
duration models described in Section~\ref{model}, which we then
compound with an estimated Gaussian distribution for trade-time
returns to synthesize a distribution for clock-time returns. Using
Monte Carlo approximations for the distribution of clock-time returns,
we evaluate the candidate models using several measures of
goodness-of-fit and distributional distance.

The choice of trade time scale, $m$, for which we estimate the
parameters of the component distributions is not a priori inherent to
the model -- in theory we could estimate the component distributions
for a variety of values for $m$. Instead, we focus exclusively
on the finest possible trade time scale, $m=1$, since this most
closely mimics the underlying trading process. With the model
estimated at such a fine time scale, we can then obtain simulations
from the distributions of clock-time returns for any clock-time
interval which is of longer duration. This would not be possible if we
estimated the trade time components for larger $m$: we would not be
able to simulate clock-time returns for clock time scales that are of
finer resolution than the chosen $m$. Although we do not report the
results here, we have estimated the model for $m > 1$ and compared
with clock-time simulations of comparable resolutions -- the results
that we report below are robust to the choice of $m$.

\subsection{Poisson/Exponential Estimation}
\label{expEst}

The assumption of Poisson-distributed trade arrivals with mean
$\gamma$ corresponds to duration times that are Exponentially
distributed with mean $\nu = \frac{1}{\gamma}$. It is trivial to show
that the maximum likelihood estimate of the Exponential mean is
\begin{gather}
  \hat{\nu} = \frac{1}{\hat{\gamma}} = \frac{1}{n} \sum_{i=1}^n
  d_i, \label{expMean}
\end{gather}
where $d_i$, $i=1,2,\ldots,n$ are the observed durations for a
given trade time scale, $m$.

Columns 1 and 2 of Table~\ref{estimates} report estimates of
$\eta$ and $\gamma$ for $m = 1$ along with parametric bootstrap
estimates of their standard errors. Although estimates of asymptotic
standard errors are readily available for these parameters (as well as
the parameters of the Gaussian distribution in
Section~\ref{gaussEst}), we report bootstrap estimates for consistency
with the MSMD results below. The reported standard errors are almost
identical to their estimated asymptotic counterparts.
\begin{table}[h]
\vspace{.1in}
\begin{center}
\scalebox{0.75}{
\begin{tabular}{cc|c|cc}
\hline 
\multicolumn{2}{c|}{Exponential/Poisson} & \multicolumn{1}{|c}{TMSMD}
& \multicolumn{2}{|c}{Tick-time Gaussian} \\
\hline 
$\nu$ & $\gamma$ & $\nu_{max}$ & $\mu$ & $\sigma$ \\
\hline 
300.7 & 3.326e-03 & 5866 & -7.103e-05 & 0.1196 \\
(2.405e-03) & (2.642e-08) & NA & (2.928e-04) & (2.072e-04) \\
\hline 
\end{tabular}
}
\end{center}
\caption{Estimates of Exponential, Gaussian and TMSMD parameters. For
  the TMSMD model, $\bar{k}= 7$ and the remaining MSMD estimates
  correspond to those given in Table~\ref{msmdEstimates}.}
\label{estimates}
\end{table}

It is important to correctly interpret the intensity parameter in
Table~\ref{estimates}. Since we are now reverting to the
assumption that the number of trades arriving in a particular time
interval, $\tau$, is distributed as a Poisson, $N_m(\tau)$ follows a
Poisson process with intensity parameter $\gamma \tau$:
\begin{gather}
  N_m(\tau) \sim Poisson(\gamma \tau). \label{poisProc}
\end{gather}
Equation~\eqref{poisProc} says that the number of $m$th trades
arriving in an interval, $\tau$, is $Poisson(\gamma \tau)$. In our
case, we estimated the Poisson parameter by fixing $m=1$ and a unit
time interval as $\tau = 1 $ millisecond. The resulting estimate
$\hat{\gamma} = 0.003326$ tells us that we expect roughly 0.003326
trades to arrive each millisecond, or roughly three trades per
second. However, given how the Poisson rate parameter scales with time
units, this also means that we expect approximately 15 trades in five
seconds, 30 trades in ten seconds and so on.

\subsection{MSMD Estimation}
\label{msmdEst}

Following \citet{Chen2013}, we evaluate the likelihood of the MSMD
model, associated with Equations~\eqref{msmd1} -- \eqref{msmd6}, using
the nonlinear filtering method of \citet{Hamilton1989} and maximize
the likelihood with a standard hill-climbing algorithm. To estimate
all parameters of the MSMD model, we iterate over candidate
values of $\bar{k}$ and estimate the remaining four parameters,
$\lambda$, $\gamma_{\bar{k}}$, $b$, and
$m_0$. Table~\ref{msmdEstimates} reports these estimates and the value
of the log likelihood at the MLEs for $\bar{k} = \{0,1,\ldots,9\}$. We
conclude that the likelihood is maximized at $\bar{k} = 3$
and declines very slowly for higher values of $\bar{k}$. In our
subsequent tests of the model (see Section~\ref{compEst}), we find
that the MSMD model with $\bar{k} = 7$ provides a slightly improved
fit to the data. This is attributed to the fact that higher values of
$\bar{k}$ result in more persistence of the duration process, which
subsequently results in greater autocorrelation in the volatility of
clock-time returns for $\tau > 5000$ ms. Otherwise, the
model results and statistical tests are very similar for $\bar{k} \in
\{3,4,5,6,7\}$. For purposes of concision, we adopt $\bar{k} = 7$
throughout the remainder of this section. The last row of
Table~\ref{msmdEstimates} reports parametric bootstrap standard errors
(using 1000 iterations) for the $\bar{k} = 7$ estimates.
\begin{table}[h]
\vspace{.1in}
\begin{center}
\scalebox{0.75}{
\begin{tabular}{cccccc}
\hline 
\multicolumn{6}{c}{MSMD} \\
\hline 
$\bar{k}$ & $\lambda$ & $\gamma_{\bar{k}}$ & $b$ & $m_0$ & Log
likelihood \\
\hline 
1 &  0.1045 & 0.5922 & 3.641 &  0.1259 & -1360966 \\
2 &  0.1039 & 0.4819 & 2.286 & 0.08913 &  -948840 \\
3 & 0.09155 & 0.4656 & 2.063 &  0.1502 &  -940237 \\
4 & 0.04678 & 0.5859 & 4.349 &  0.1395 &  -941637 \\
5 &  0.3338 & 0.5815 & 4.269 &  0.1395 &  -941641 \\
6 &  0.1797 & 0.5870 & 4.423 &  0.1388 &  -941687 \\
7 & 0.09660 & 0.5884 & 4.461 &  0.1386 &  -941698 \\
8 & 0.05190 & 0.5887 & 4.471 &  0.1386 &  -941700 \\
9 &  0.3743 & 0.5883 & 4.460 &  0.1386 &  -941698 \\
\hline 
Std. Errors ($\bar{k}=7$) & (1.314e-02) & (3.962e-03) & (4.801e-02) &
(3.704e-04) \\
\hline 
\end{tabular}
}
\end{center}
\caption{Estimates and log likelihood values of MSMD models for
  $\bar{k}= \{1,\ldots,9\}$.}
\label{msmdEstimates}
\end{table}

\begin{figure}[ht]
\begin{center}
\includegraphics[width=6.25in]{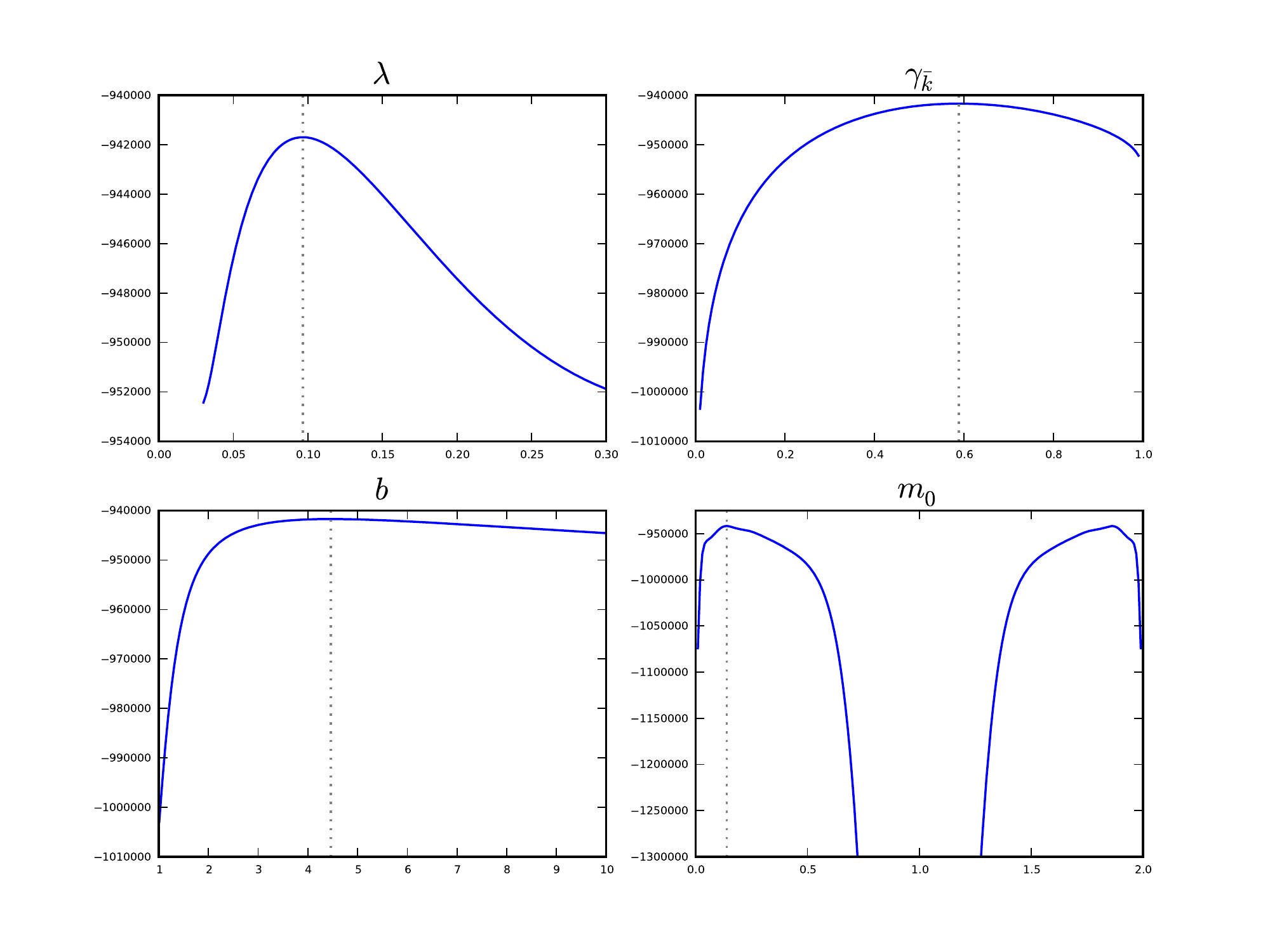}
\caption{Profiles of the MSMD log-likelihood surface at the achieved
  maximum (for $\bar{k}=7$) in the coordinate directions of the model
  parameters.}
\label{likeMap}
\end{center}
\end{figure}

Figure~\ref{likeMap} depicts profiles of the log likelihood in the
coordinate directions of the $\bar{k}=7$ parameter estimates reported
in Table~\ref{msmdEstimates}.
The profiles show that the log likelihood is smooth
and well behaved in the vicinity of the estimates. One interesting
feature that we note is the symmetry of the log likelihood with
respect to $m_0$ -- this is a direct result of the fact that the
random variable $M$ takes the values $m_0$ and $2-m_0$ with equal
probability. As noted by \citet{Chen2013}, the parameters
$\gamma_{\bar{k}}$ and $b$ are not identified when $m_0=1$, which is
manifest as asymptotic behavior in the log-likelihood profile. As a
result, two values of $m_0$ achieve the maximum and it is irrelevant
which we choose. In our application, we have chosen $\hat{m}_0 \in
(0,1)$.

\subsection{TMSMD Estimation}
\label{tmsmdEst}

Following Section~\ref{compTMF}, the truncated MSMD model is
formulated by independently combining the MSMD model with an
Exponential distribution that is selected to fit the maximal
inter-trade duration observed in the data. Durations under the
TMSMD model are simply the minimum of durations generated by the
independent components. For this reason, the estimates of the MSMD
component are identical to those reported above for the stand-alone
MSMD model, again with $\bar{k}=7$.

Our objective in selecting an Exponential distribution for the TMSMD
model is to find a random variable, $d_{\overline{Exp}}$, such that
\begin{gather*}
  d_{\overline{Exp}} \sim Exp(\nu_{max}) \\
  \E\left[\max\{d_{\overline{Exp}}\}\right] = d_{max},
\end{gather*}
where $d_{max} = 56315$ ms is the longest inter-trade duration
observed in our sample of passive-period E-mini data.

Appendix~\ref{extremum} derives the expectation of
$\max\{d_{\overline{Exp}}\}$:
\begin{align}
  \E\left[\max\{d_{\overline{Exp}}\}\right] & = \nu_{max} \sum_{i=1}^n
  \frac{1}{i}. \label{maxdmaxExp}
\end{align}
We note that the expected value depends on the number of observed
durations, which is implicitly a function of the expected duration,
$\nu_{max}$, itself. As an approximation, we set $n$ equal to the
total time in our data sample (48,600,000 milliseconds) divided by the
average duration and rounded to the nearest integer:
\begin{align*}
  n & = \left[\frac{48,600,000}{\nu_{max}}\right].
\end{align*}
Thus, to calibrate an appropriate value for $\nu_{max}$ we choose
$\hat{\nu}_{max}$ to minimize the function
\begin{align*}
  \hat{\nu}_{max} & = \underset{\nu}{\text{argmin}} \left\{\left(\nu
  \sum_{i=1}^{[48600000/\nu]} \frac{1}{i} - 56315\right)^2
  \right\}.
\end{align*}
The numerical solution to this problem is $\hat{\nu}_{max} = 5866$ and
is reported in Table~\ref{estimates}. We do not provide a standard
error for this parameter since it is computed via numerical
optimization using observed data and is highly sensitive to the total
duration time and maximal value of the data. Estimating the standard
error via parametric bootstrap would be unreliable due to variation
within simulated durations under any candidate model. Truncating the
MSMD model of the previous section with this Exponential provides a
much better fit to both the tail of the distribution of observed
durations as well as their sample autocorrelations, as shown in
Section~\ref{comparison}.

\subsection{Trade-time Gaussian Estimation}
\label{gaussEst}

The key observation of this paper, highlighted in
Section~\ref{distributions}, is that trade-time returns observed
outside of pre-scheduled news periods are well characterized
by a Gaussian distribution. Furthermore, the near lack of serial
correlation exhibited by the returns and corresponding squared returns
indicates that they can be modeled as independent and identically
distributed. In this case, the maximum likelihood estimates of the
parameters of the distribution are simply the sample average and
standard deviation of trade-time returns, measured for a specific
value of $m$. The last two columns of Table~\ref{estimates} report
these estimates for $m=1$. The corresponding bootstrap standard errors
are reported in parentheses below the estimates.

\subsection{Simulating Clock-time Returns}
\label{compEst}

With estimates of the component distributions in hand, we obtain 
Monte Carlo approximations of the clock-time returns distribution
using the Gaussian mixture model, expressed in
Equation~\eqref{mix}. We do this in a hierarchical fashion, first
simulating inter-trade durations for $m=1$ from the Exponential, MSMD
and TMSMD models, pairing the durations with independent draws of
trade-time returns from the estimated Gaussian density, and finally
aggregating individual returns within a fixed clock time interval.

Following the procedure outlined above, we aggregate returns for
clock-time intervals $\tau = \{250,\allowbreak 500,\allowbreak
1000,\allowbreak 5000,\allowbreak 10000,\allowbreak 30000\}$
milliseconds until we obtain $n=\{208000,\allowbreak
104000,\allowbreak 52000,\allowbreak 10400,\allowbreak
5200,\allowbreak 1716\}$ clock-time returns, respectively, which
correspond to the number of observations in the data for those
time intervals. The individual simulations under the Exponential, MSMD
and TMSMD models use the same trade-time returns; they only differ in
the elapsed time between observations. It is important to mention
three adjustments that we make in order to simulate clock time
returns. First, since E-mini returns are discrete and only observed at
increments of 0.25 points, we simulate tick-time returns from the
continuous Gaussian distribution described above and then discretize
to the nearest 0.25 increment. For example, a simulated tick-time
return of 0.13 would be discretized to 0.25, while a simulated
tick-time return of 0.12 would be discretized to zero. Second, we
perform a similar discretization of simulated durations (under all
models) by rounding values to the nearest millisecond. Since zero
durations are not allowed in our framework, all simulated durations
below one millisecond  are rounded upward. Finally, the statistical
tests we report below require that the models place probability mass
(for the discretized distributions of clock-time returns) on the same
support as the empirical density of the data. In some cases, however,
aggregated simulated returns under the models are observed
outside the set of values observed in the data. In those cases we
simply set the returns to zero. While this has the potential to impact
the results of the model, in all but one case (noted below) we revise
so few of the returns that the adjustment is of little empirical
relevance. Table~\ref{threshhold} reports the number of values (and
their fraction of the total simulation) that were adjusted in this
manner, for each model and for each time-scale. The numbers are quite
low for fine time scales for all models and increase with $\tau$. The
stand-alone MSMD model performs best in this dimension, with
virtually no adjustment, while the Exponential model adjusts 16\% of
returns when $\tau = 30000$ ms. For smaller values of $\tau$, however,
the Exponential model produces substantially fewer clock-time returns
outside the empirical support. Finally, in the worst case ($\tau =
30000$ ms) the TMSMD model requires an adjustment of 1.1\% of
simulated returns, but for all other values of $\tau$ the fraction is
well below 1\%, often by one or two orders of magnitude.
\begin{table}[h]
\vspace{.1in}
\begin{center}
\scalebox{0.75}{
\begin{tabu}{lcccccc}
\hline 
& \multicolumn{6}{c}{$\tau$} \\
\cline{2-7} 
& 250 & 500 & 1000 & 5000 & 10000 & 30000 \\
\hline 
\rowfont{\color{expCol}} Exp counts (fraction) & 2 (1e-05) & 10
(0.0001) & 63 (0.00126) & 158 (0.0158) & 149 (0.0298) & 272 (0.16) \\
\rowfont{\color{msmdCol}} MSMD counts (fraction) & 0 (0.0) & 1 (1e-05)
& 1 (2e-05) & 1.0 (0.0001) & 0 (0.0) & 0 (0.0) \\
\rowfont{\color{tmsmdCol}} TMSMD counts (fraction) & 35 (0.000175) &
40 (0.0004) & 47 (0.00094) & 24 (0.0024) & 15 (0.003) & 19 (0.011176)
\\
\hline 
\end{tabu}
}
\end{center}
\caption{Number (fraction in parentheses) of simulated clock-time
  returns that were adjusted to fall within the support of discretely
  observed data values.}
\label{threshhold}
\end{table}

\begin{figure}[ht]
\begin{center}
\includegraphics[width=6.5in]{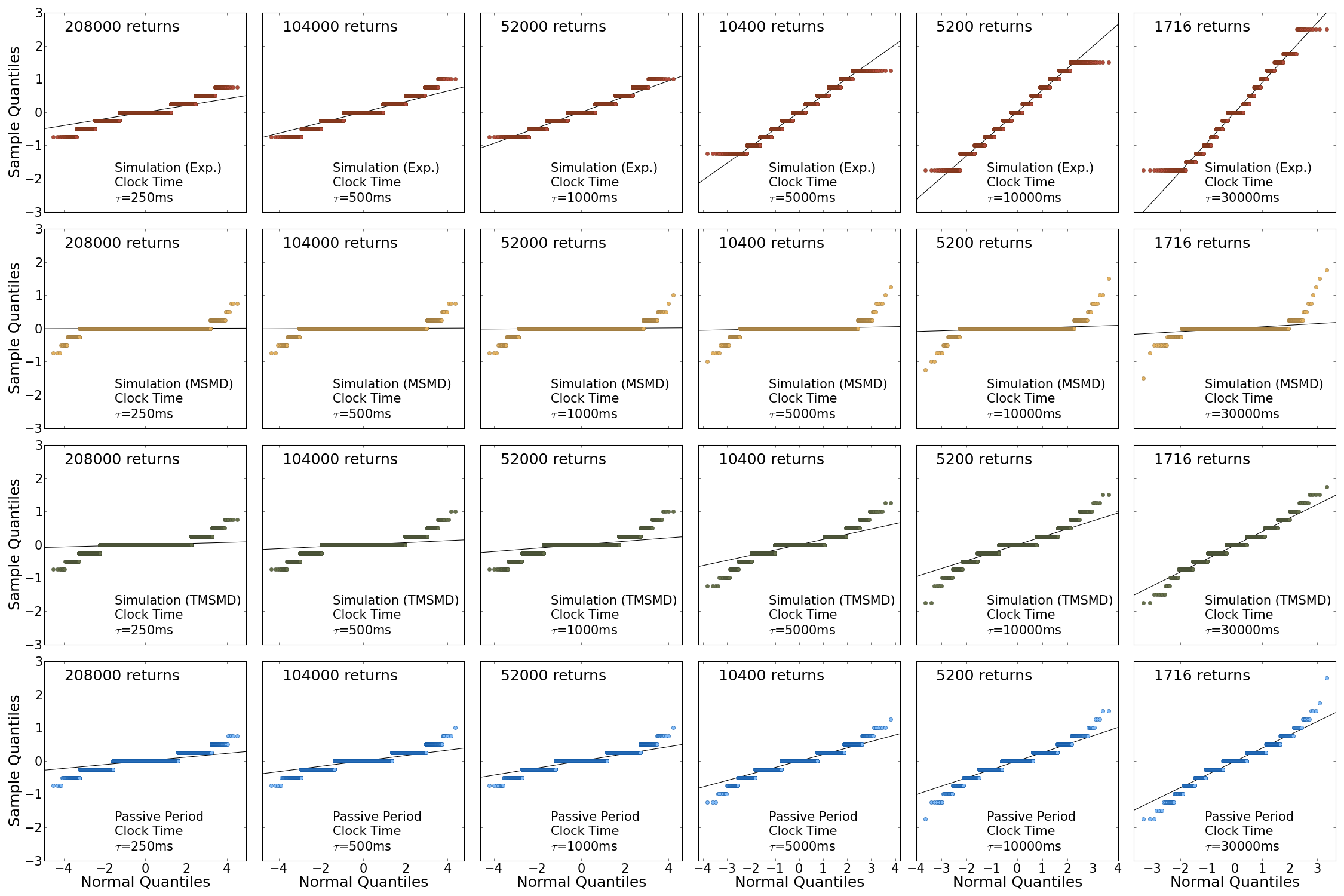}
\caption{Sample Q-Q plots for simulated clock-time returns. The upper
  three rows of panels correspond to returns simulated under the
  Exponential, MSMD and TMSMD models (respectively), for several clock
  time intervals $\tau$. The bottom row of panels is a reproduction
  of the E-mini passive-period clock-time Q-Q plots show in
  Figure~\ref{qqAll}.}
\label{modelQQ}
\end{center}
\end{figure}

Figure~\ref{modelQQ} shows Q-Q plots of the clock-time returns
simulations for each value of $\tau$ that we consider. The panels in
the first three rows correspond to the Exponential, MSMD and TMSMD
models, respectively. The panels in the final row of the figure are a
reproduction of the E-mini passive-period clock-time Q-Q plots shown
in Figure~\ref{qqAll}. It is immediately apparent from the plot that
clock-time returns under the MSMD and TMSMD models exhibit heavy tails
for all values of $\tau$ while the Exponential model does a very poor
job of capturing leptokurtosis, except for the lowest values of
$\tau$. This is attributed to the particular nature of the MSMD model:
it can be interpreted as a mixture of Exponential distributions, which
does a good job of capturing over-dispersion of observed inter-trade
durations relative to a simple Exponential. In particular, the
persistence of the latent states in the MSMD model generates more
variation in inter-trade durations relative to the Exponential
distribution, which leads to a more heterogeneous mixture of Gaussian
densities in Equation~\eqref{mix}, resulting in a greater degree of
leptokurtosis.

It is also immediately apparent from Figure~\ref{modelQQ} that the
distributions of TMSMD returns are much more similar to the data than
those of MSMD returns. In fact, the MSMD returns distributions
allocate an excessive portion of probability mass to $r_{\tau}(t) =
0$. This is directly attributable to the distribution of MSMD
durations, which is depicted in Figures~\ref{durHist} and
\ref{durQQ}. As seen in those figures, the MSMD distribution has a
much heavier right tail than the distribution of the data. This
results in a large fraction of trades separated by very long
durations, which causes prices to remain constant (zero returns)
across clock-time intervals with a frequency that is too high. The
TMSMD model rectifies this problem by truncating the long right tail
of the MSMD distribution and more closely fitting the tail of the
empirical distribution observed in the data. The result is that returns
under the TMSMD model conform much more closely to those of the data.

To provide a formal measure of fit, the first three rows of
Table~\ref{gofStats} report Chi-squared test statistics for returns
distributions under each of the three duration models, relative to the
empirical distribution of returns. The Chi-squared test is a test of
similarity for discrete distributions: under the null
hypothesis of identical distributions, a properly weighted sum of the
discrepancies in the histograms of each model and the data should be
distributed as a $\chi^2(k)$, where the degrees of freedom $k$ is one
less than the number of histogram bins (unique values that the random
variable can assume). Quantiles corresponding to probability 0.95 are
reported in the fourth row of Table~\ref{gofStats} -- Chi-squared test
statistics that exceed these values represent a rejection of the null
hypothesis at the 5\% level.
\begin{table}[h]
\vspace{.1in}
\begin{center}
\begin{tabu}{lcccccc}
\hline 
& \multicolumn{6}{c}{$\tau$} \\
\cline{2-7} 
& 250 & 500 & 1000 & 5000 & 10000 & 30000 \\
\hline 
\rowfont{\color{expCol}} $\chi^2$ Exp & 35035.0  & 63146.0 & 106900.0 & 77396.0 & 48035.0 & 8680.0 \\
\rowfont{\color{msmdCol}} $\chi^2$ MSMD & 24747.0  & 20079.0 &  15229.0 &  7519.8 &  5252.7 & 3039.9 \\
\rowfont{\color{tmsmdCol}} $\chi^2$ TMSMD & 15462.0  & 10963.0 &   6577.7 &  762.14 &  85.534 &  50.68 \\
5\% $\chi^2_k$ critical values & 12.592 & 14.067 & 14.067 & 18.307 & 22.362 & 24.996 \\
\hline 
\rowfont{\color{expCol}} KL Exp & 0.032975 & 0.089006 & 0.20827 &  0.50814 &   0.55751 &   0.4007 \\
\rowfont{\color{msmdCol}} KL MSMD &  0.43104 &  0.62514 & 0.81613 &   1.2171 &    1.2877 &    1.365 \\
\rowfont{\color{tmsmdCol}} KL TMSMD & 0.089029 &  0.11033 & 0.11152 & 0.041454 & 0.0076904 & 0.014179 \\
\hline 
\end{tabu}
\end{center}
\caption{Goodness of fit statistics for simulated returns under the
  Exponential and MSMD duration models. The first three rows report
  $\chi^2$ goodness of fit measures relative to the observed data for
  each clock time interval $\tau$ that we consider. The latter three
  rows report Kullback-Leibler divergences, again, relative to the
  observed data.}
\label{gofStats}
\end{table}
The table clearly demonstrates that each of the models fails the
Chi-squared goodness-of-fit test at the 5\% level for all values of
$\tau$. However, this result is not surprising: we know \emph{a
  priori} that none of our models is an \emph{exact} characterization
of the true data generating process. Our objective, instead, is to
find a model that is a suitable approximation. In this case, with an
extremely large dataset, the Chi-squared test is simply informing us
that we have a lot of data and that our models are not \emph{exactly}
correct. What is more interesting, however, are the magnitudes of the
Chi-squared statistics: the MSMD statistics are uniformly better than
those of the Exponential (often by an order of magnitude) and the
TMSMD statistics are uniformly better than those of the MSMD (by as
much as two orders of magnitude). In fact, while the TMSMD model
is rejected for $\tau = 10000$ and $\tau=30000$ ms, the Chi-squared
statistics are remarkably close to their empirical counterparts when
considering the size of the data sample and the simulation.

To provide a another measure of distributional distance, the final
three rows of Table~\ref{gofStats} report the Kullback-Leibler
divergence for the returns distributions under each of the three
duration models, relative to the empirical distribution of
returns. In contrast to the Chi-squared test, the Kullback-Leibler
divergence (see \citet{Kullback1951}) is not a statistical test of a
formal hypothesis -- rather, it is a measure of information loss when
using one distribution as an approximation for another. For discrete
distributions $F$ and $G$ over values $\{x\}_{i=1}^n$, the
Kullback-Leibler divergence of $G$ from $F$ is defined as
\begin{align*}
  D(F||G) & = \sum_{i=1}^n \log\left(\frac{F(x_i)}{G(x_i)}\right)
  F(x_i),
\end{align*}
which is the expectation (under distribution $F$) of log probability
ratios. In contrast to the magnitudes of the Chi-squared statistics,
Table~\ref{gofStats} shows that the distribution of returns under the
Exponential model is closest to that of the data for $\tau = 250$ ms
and $\tau = 500$ ms and is uniformly closer than the distribution of
MSMD returns. However, for larger values of $\tau$, the TMSMD
model dominates, often by one or two orders of magnitude. The reason
for the failure of the MSMD model under this metric is the high
probability mass placed on $r_{\tau}(t) = 0$.

\begin{figure}[ht]
\begin{center}
\includegraphics[width=6.5in]{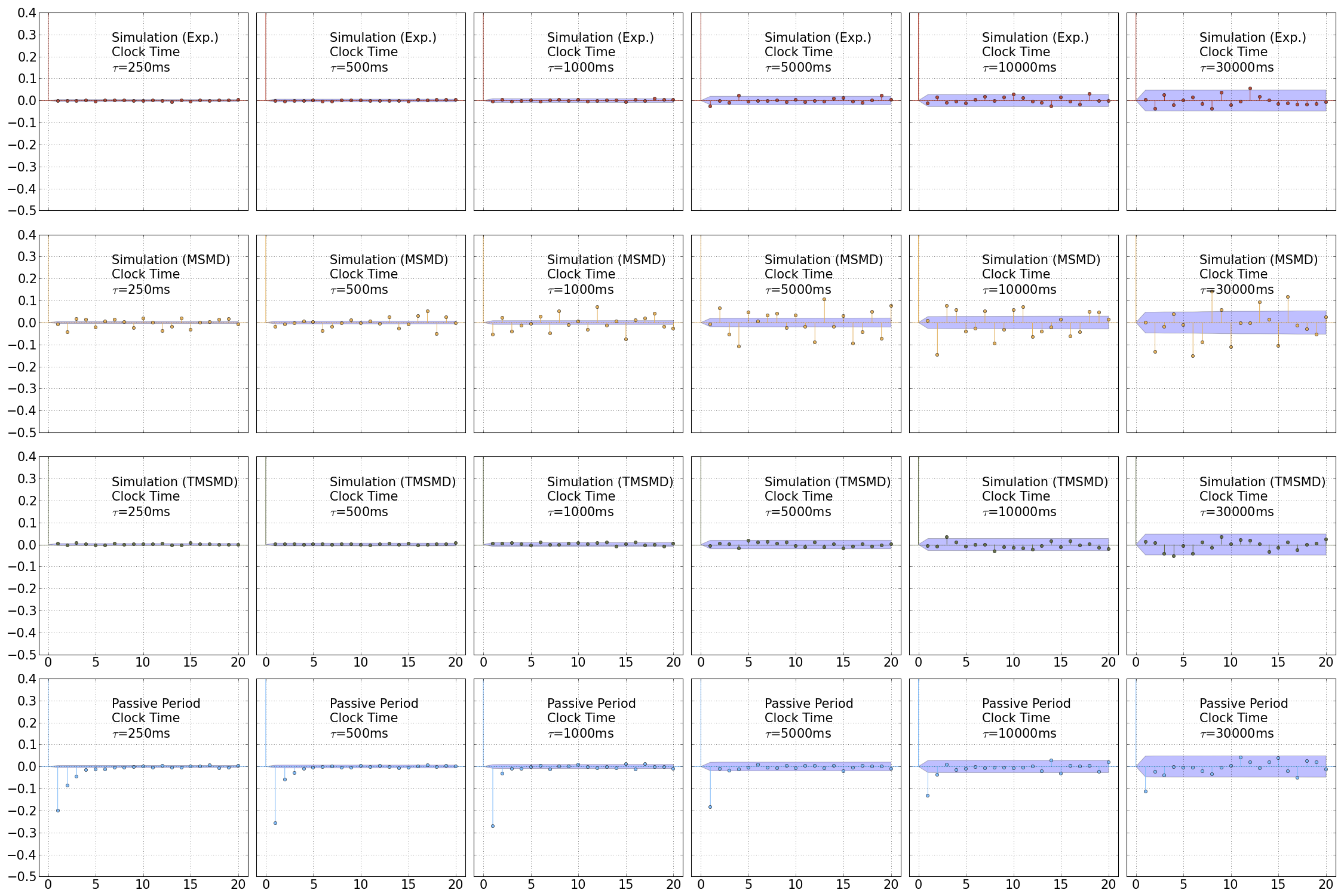}
\caption{Sample autocorrelation functions for simulated clock-time
  returns. The upper three rows of panels correspond to returns
  simulated under the Exponential, MSMD and TMSMD models, for several
  clock time intervals $\tau$. The bottom row of panels is a
  reproduction of the autocorrelations of E-mini passive-period
  clock-time returns in Figure~\ref{acAll}.}
\label{modelAC}
\end{center}
\end{figure}

Figure~\ref{modelAC} shows sample autocorrelation functions for returns
simulated under each of the duration models, with each column of
panels corresponding to a time scale $\tau = \{250,\allowbreak 500,\allowbreak 1000,\allowbreak
5000,\allowbreak 10000,\allowbreak 30000\}$ milliseconds. As with
Figure~\ref{modelQQ} the first three rows depict autocorrelations under
the Exponential, MSMD and TMSMD models, respectively, while the last
row is a reproduction of the autocorrelations of E-mini passive-period
clock-time returns in Figure~\ref{acAll}. Much like the data, the
returns ACFs of the models exhibit little autocorrelation,
although the MSMD model appears to have a frequency of non-zero
autocorrelations that is too high and which exhibit no pattern. While
none of the models captures the negative autocorrelation attributed to
bid/offer bounce and mean reversion at low lags in finely sampled
data (low $\tau$), this dynamic is not explicitly modeled in our
framework and is not expected to be present.

\begin{figure}[ht]
\begin{center}
\includegraphics[width=6.5in]{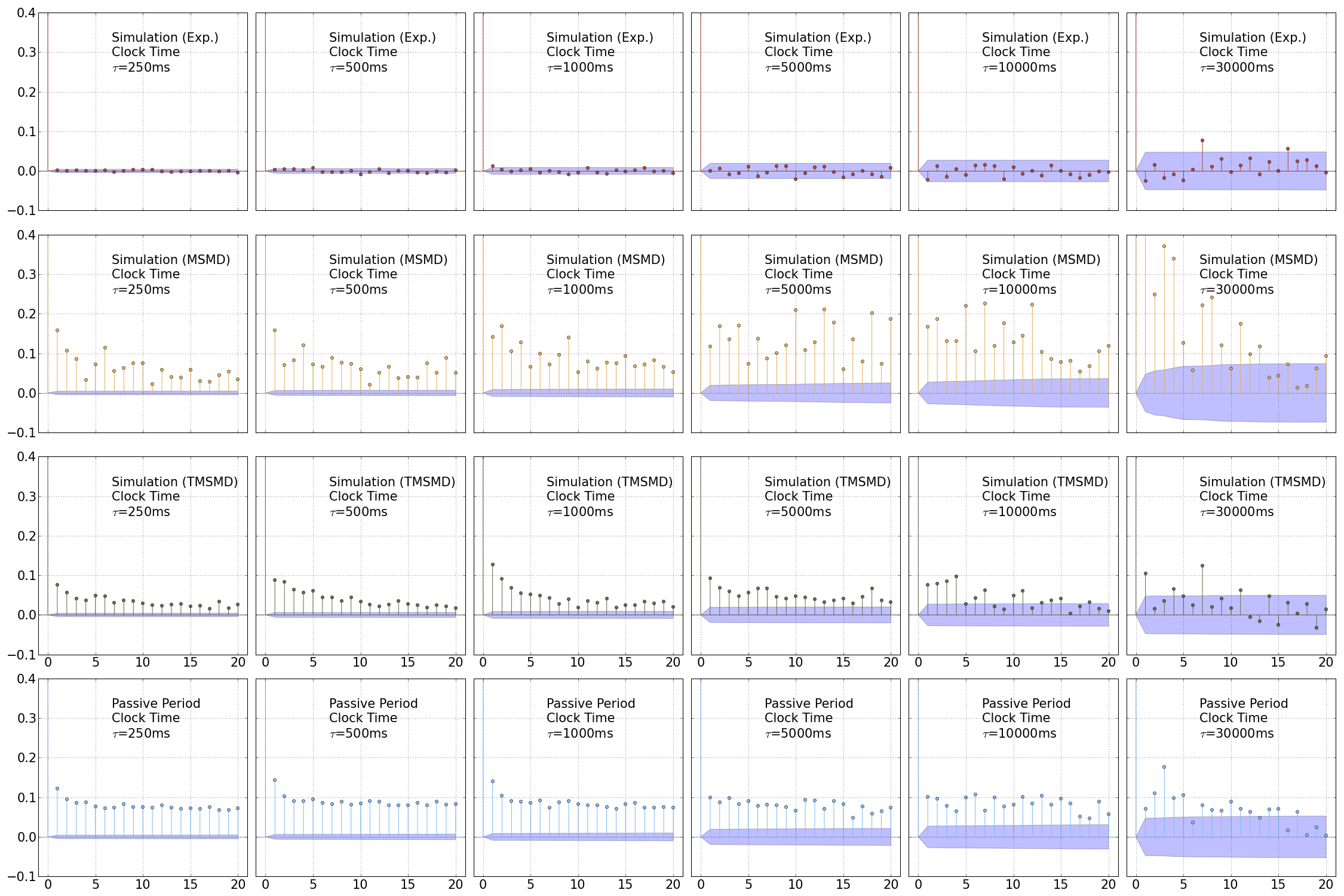}
\caption{Sample autocorrelation functions for simulated clock-time
  \emph{squared} returns. The upper three rows of panels correspond to
  returns simulated under the Exponential, MSMD and TMSMD models, for
  several clock time intervals $\tau$. The bottom row of panels is a
  reproduction of the autocorrelations of E-mini passive-period
  clock-time squared returns in Figure~\ref{ac2All}.}
\label{modelAC2}
\end{center}
\end{figure}

Sample autocorrelations of squared returns simulated under each of the
models are shown in Figure~\ref{modelAC2}. Persistence among the
autocorrelations is present in the MSMD and TMSMD models but not in
the Exponential. More importantly, the TMSMD autocorrelations of
squared returns conform much more to the data than those of the
stand-alone MSMD. This dynamic is a great strength of the framework
we promote: the truncated compound multifractal process can jointly
explain leptokurtosis and volatility clustering. In particular,
Figure~\ref{modelAC2} shows that autocorrelations of squared returns
under the TMSMD model are generally smaller than those of the
passive-period E-mini data, but that they exhibit similar decline
with increasing time scale, $\tau$.

As a final measure of model fit, Table~\ref{lbStats} reports Ljung-Box
statistics for the autocorrelation functions of returns and squared
returns, both in the data and under each of the three models. The
Ljung-Box statistic is defined as
\begin{align*}
  Q & = n(n+2) \sum_{i=1}^l \frac{\hat{\rho}^2_i}{n-i},
\end{align*}
where $\hat{\rho}_i$ is the sample autocorrelation for lag $i$, $n$ is
the number of observations in the data, and $l$ is the number of lags
over which the statistic is computed. Under the null hypothesis that
all autocorrelations are jointly zero, $Q \sim \chi^2(l)$. For the
statistics reported in Table~\ref{lbStats} we set $l=20$, but the
results were robust to a variety of other choices. This choice of $l$
dictated a common 5\% critical value of $\chi^2_{0.95}(20) = 31.41$.
\begin{table}[h]
\vspace{.1in}
\begin{center}
\begin{tabu}{lcccccc}
\hline 
& \multicolumn{6}{c}{$\tau$} \\
\cline{2-7} 
& 250 & 500 & 1000 & 5000 & 10000 & 30000 \\
\hline 
\rowfont{\color{expCol}} LB $r_{Exp}$       &    25.2 &  28.707 &  25.64 &  19.6  & 18.261 & 18.848 \\
\rowfont{\color{msmdCol}} LB $r_{MSMD}$      &  860.02 &  483.01 & 311.56 & 95.955 & 95.207 & 76.485 \\
\rowfont{\color{tmsmdCol}} LB $r_{TMSMD}$   &  45.206 &  53.868 & 47.937 & 23.578 & 20.614 & 28.109 \\
\rowfont{\color{passiveCol}} LB $r_{Data}$      &  1930.1 &  351.88 & 110.36 & 18.059 & 28.569 & 20.303 \\
\hline 
\rowfont{\color{expCol}} LB $r_{Exp}^2$     &  32.536 &  11.004 & 22.702 & 16.973 &  19.24 & 16.647 \\
\rowfont{\color{msmdCol}} LB $r_{MSMD}^2$    &  2397.5 &  2435.3 & 1855.0 &  409.8 & 452.78 & 135.69 \\
\rowfont{\color{tmsmdCol}} LB $r_{TMSMD}^2$ &  6559.7 &  4135.7 & 1780.7 & 624.93 &  345.7 & 68.076 \\
\rowfont{\color{passiveCol}} LB $r_{Data}^2$    & 27412.0 & 16306.0 & 6767.5 & 1249.4 & 625.47 & 190.64 \\
\hline 
\end{tabu}
\end{center}
\caption{Ljung-Box statistics for simulated returns under the
  Exponential, MSMD and TMSMD duration models as well as the observed
  data for each clock time interval $\tau$ that we consider. The first
  four rows report Ljung-Box statistics for the ACFs of returns and
  the last four rows report Ljung-Box statistics for the ACFs of
  squared returns.}
\label{lbStats}
\end{table}
The upper four rows of the Table~\ref{lbStats} report Ljung-Box
statistics for the ACFs of returns. Interestingly, the data do not
pass the test at the 5\% level for the lowest three values of $\tau$,
due to the large negative autocorrelation attributed to bid/offer
bounce and mean reversion. In contrast, the Exponential returns never
fail the test and the MSMD returns always fail the test, although
their test statistics decline for increasing $\tau$, similar to the
data. TMSMD returns fail the test for precisely the same values of
$\tau$ as the data, but visual inspection of Figure~\ref{modelAC}
does not suggest that there is a systematic reason for this. However,
for $\tau \geq 5000$ ms, TMSMD returns do not fail the Ljung-Box test
and have test statistics that are very similar in magnitude to the
data.

The lower four rows of Table~\ref{lbStats} report Ljung-Box
statistics for the ACFs of \emph{squared} returns. Both the MSMD and
TMSMD models mimic the data and fail the test for all values of $\tau$
while the Exponential model never fails. This is a confirmation of the
inability of the compound Poisson process to explain volatility
clustering. While the magnitude of the MSMD statistics (for squared
returns) are closer to those of the data for $\tau = 10000$ ms and
$\tau = 30000$ ms, the TMSMD statistics follow the data a bit better.

In sum, we find that the truncated compound multifractal model does a
good job of capturing both fat tails and volatility persistence in
clock time, while leaving the level of returns with almost no
autocorrelation. These features are primarily attributed to the
underlying MSMD model. The compound Poisson process (corresponding to
durations drawn from an Exponential distribution), in contrast,
generates clock-time returns that are serially uncorrelated but which
do not exhibit volatility persistence nor the fat tails observed in
actual data.

\section{Application: Price Formation and Equity Market Structure}
\label{application}

During trading periods that are not directly associated with news
events, the foregoing compound TMSMD/Gaussian model provides a direct
connection between the trading rate and the realized volatility of a
given security. We now utilize this connection to predict market
behavior during periods of stress.

We illustrate this connection with additional historical trading data
for the CME E-mini near-month S\&P500 futures contract. With the use
of  the near-month E-mini contract, we are drawing on the same
instrument that we used to build our model. We elect, however, to
examine a different (and substantially longer) historical range, so
that the data is out-of-sample with respect to that used to construct
the compound TMSMD/Gaussian model. Specifically, the new data set
contains
millisecond-stamped trade events for 577 trading sessions between
April 27th, 2010 and August 17th 2012. This data spans periods that
exhibited substantially higher volatility (such as the days
surrounding the Flash Crash in early May 2010, and the market downturn
of August 2011) than occurred during our May -- August, 2013 in-sample
period.

For each of the 577 days in the data set, we collect
millisecond-precision time stamps for all E-mini trades occurring
between 3:00 p.m. and 4:00 p.m. Eastern Time. This interval
corresponds to the last hour of the U.S. equity market trading
day. The trading rate is generally high during the last hour of the
equity market session and there are very few scheduled market-moving
news releases during this daily window. As before, when multiple
trades occur during a single millisecond, we aggregate them as a
single transaction and use the final, in-force, price of the
millisecond as the price of the trade. Finally, we calculate the
median inter-trade duration, $\tau_{\rm med}$, among these trades on
each day.

\begin{figure}[ht]
\centering
\includegraphics[width=0.84\textwidth]{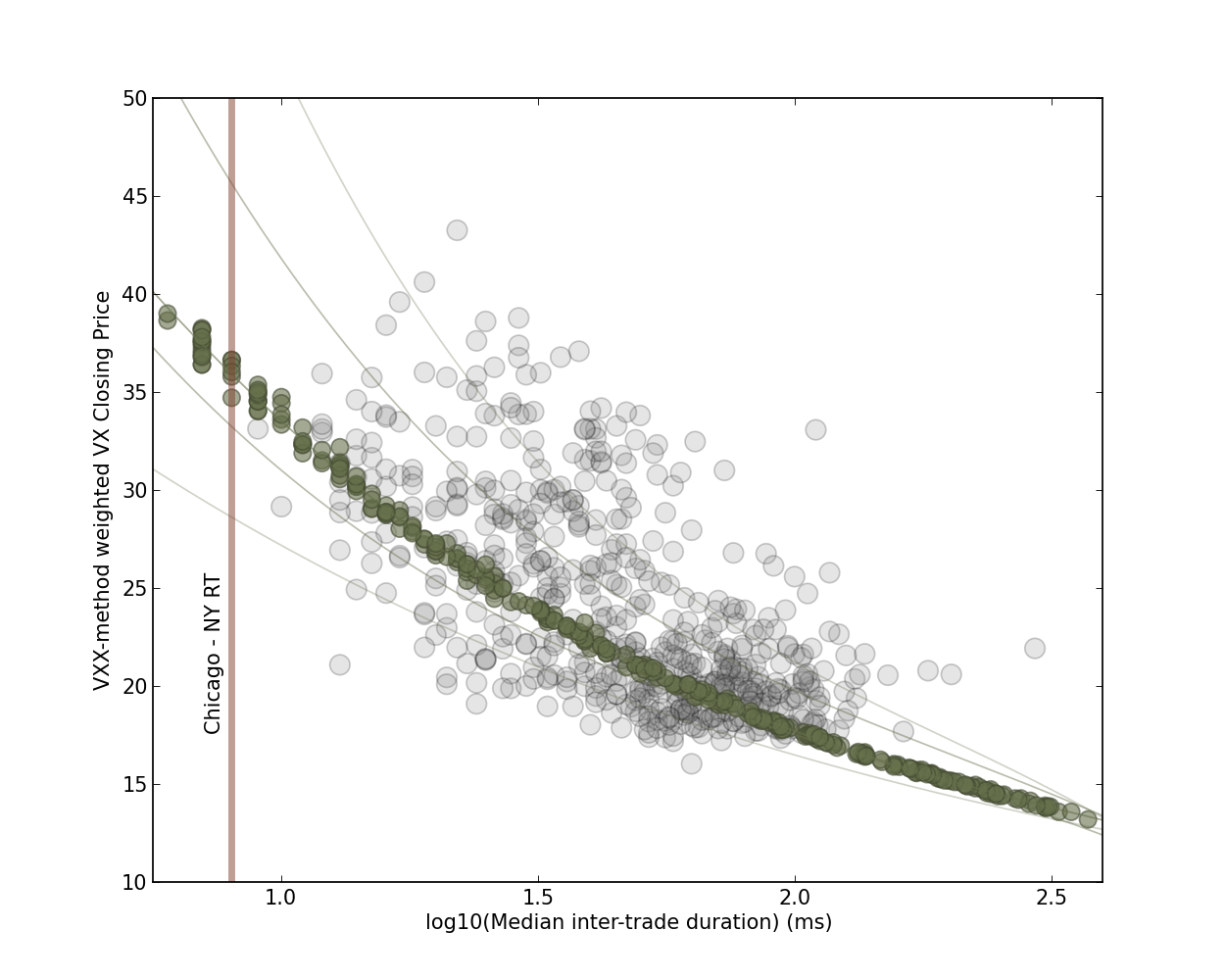}
\caption{\textit{Gray circles:} Observed daily closing prices for
  VX futures (with weightings adjusted to achieve a constant one-month
  maturity) vs. the daily median inter-trade durations observed for
  the near-month E-mini S\&P 500 futures contract trading from 3:00
  p.m. through 4:00 p.m Eastern Time. \textit{Green Circles:} Annualized
  volatilities generated from simulations of the TMSMD model
  ($\bar{k}=5$) driving a Gaussian random walk. The solid lines are
  cubic polynomial fits to simulated data under the TMSMD model with
  $\bar{k} = \{3,4,5,6,7\}$. The simulated data is only depicted for
  $\bar{k}=5$ in order to retain clarity.}
\label{fig:VXX_vs_rate}
\end{figure}

For each of the 577 sessions, we also recorded the daily closing
prices of the near-month and second-month VX-CBOE Volatility Index
Futures, drawing from data published by the
CBOE\footnote{http://cfe.cboe.com/Data/HistoricalData.aspx}. We then
combined the near- and second-month futures closing prices to create a
constant weighted-average futures maturity of one month, using the
same methodology underlying the iPath S\&P 500 VIX Short-Term Futures
ETN (ticker symbol
VXX\footnote{http://www.ipathetn.com/static/pdf/vix-prospectus.pdf}). This
computation is not an exact replication of the VXX closing price,
since it does not account for reverse splits and the daily
roll. However, it is a tradeable surrogate for the value of the VIX. The 
data are shown as gray circles in Figure \ref{fig:VXX_vs_rate}, which
presents a clear correlation between volatility futures prices and the
E-mini trading rate.

For a specific set of parameters, the compound TMSMD model permits
construction of synthetic sequences of clock-time returns and their
corresponding annualized volatilities. Using the TMSMD and Gaussian
estimates in Tables~\ref{estimates} and \ref{msmdEstimates}, we vary
the baseline intensity parameter $\lambda \in [0.01,6]$ to map out
sequences of returns with the same median durations as those observed
in the April 2010 -- August, 2012 data and then compute the
model-implied annualized volatility corresponding to each median trade
duration. These values are depicted in Figure~\ref{fig:VXX_vs_rate} as
green circles for the TMSMD model with $\bar{k}=5$, along with a cubic
polynomial that is fit to the simulated data. The remaining lines on
the figure are similar polynomials fit to simulated data under the
TMSMD model with $\bar{k} = \{3,4,6,7\}$. The uppermost line
corresponds to $\bar{k}=3$ and the bottommost to $\bar{k} =
7$. Interestingly, although the TMSMD model with $\bar{k} = 7$
provides the best in-sample fit to the May -- August, 2013 data
(primarily due to the persistence of durations), $\bar{k} = 5$
provides the best out-of-sample fit to the trade rate/volatility
relationship. In any case, our subset of models sweeps the extent of
the observed data and reproduces the general empirical
relationship. The TMSMD model also provides a basis for extrapolation
into regimes of high and low volatility that were not observed during
the 577 trading-day period.

Between April 2010 and August 2012, the smallest median inter-trade
duration during the last hour of trading occurred on June 6th, 2010,
with an observed value of $\tau_{med}=9~$ms. The
one-month maturity VX-weighted closing price on that day was $V_{\rm
  VXX}=33.12$. At realized volatilities that are only somewhat larger
than those observed in our 577 day sample, our model predicts that the
median inter-trade duration at the CME will drop substantially
\textit{below} $\tau_{\rm med}=10~$ms. To the extent that the trading
activity at the CME is the sole determinant of the E-mini futures
price, and assuming that the trading infrastructure itself is able to
function smoothly in the presence of increased messaging and
computational load, no special change in behavior would be expected
for values of $\tau_{\rm med}$ in the $1~$ms -- $10~$ms range.

Indeed, price formation in the U.S. equity market is traditionally
assumed to occur in the E-mini contract, as a consequence of the
instrument's composition, its liquidity, and its notional transacted
value (\citet{Hasbrouck2003}). The fungibility, however, of the E-mini
contract into its 500 component stocks, means that equity trading
(either in the component stocks themselves, or in tracking instruments
such as SPY (State Street Advisors S\&P 500 ETF)) may control a
non-negligible fraction of the price discovery process.

The physical separation between the CME order matching engine
(located in Aurora, Illinois) and the matching engines for the
U.S. equities exchanges (all located in suburban New Jersey) provide
an opportunity to quantitatively assess the price discovery
process. If one trading venue is responsible for price formation, then
price innovations will propagate to outlying exchanges and a clear
lead-lag relationship will be established. \citet{Laughlin2014} have
shown that price formation can be evaluated by analyzing the lag
structure of price-changing trades among highly correlated securities
-- specifically the E-mini near-month futures contract and the SPY
ETF. Recent  infrastructure improvements now routinely permit market
data and trading orders to flow between physically distant exchanges
at nearly the speed of
light\footnote{\url{http://www.wallstreetandtech.com/trading-technology/latency-at-the-speed-of-light/d/d-id/1268776?}},
which is just above 8 milliseconds round-trip between Aurora, IL and
suburban NJ.

To evaluate where price formation actually occurs, we step
sequentially through the E-mini trade records within the 577 day data
set described above. During periods when both the CME futures exchange
and the U.S. equities exchanges are open, we identify in-force
prices (the most recent traded price) at the end of each millisecond
interval and screen for the occurrence of E-mini trades in which the
in-force trade exhibits a change in price over the most recent
in-force trade from a previous 1-ms interval.\footnote{Our
  millisecond-stamped SPY trade records for  April 27th, 2010 through
  August 17th 2012 are obtained from the NYSE TAQ data for the
  U.S. Consolidated Market System covering NASDAQ, NYSE Arca, NYSE,
  BATS BZX, BATS BYX, EDGX, NASDAQ, CBSX, NSX, and CHX. During the
  period covered by the analysis, the average daily notional value of
  SPY trades was \$24.6 Billion.}

\begin{figure*}[ht]
 \centering
 \includegraphics[width=1.0\textwidth,angle=0]{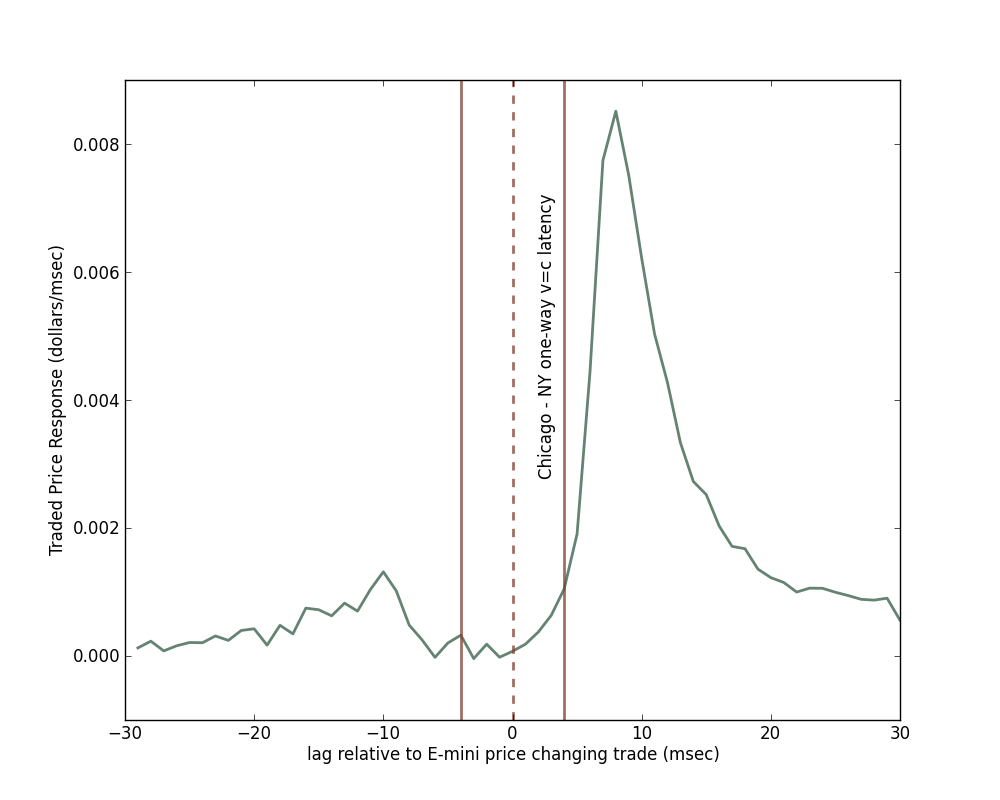}
\caption{Traded SPY price response to E-mini price-changing
  trades. The $\pm$4ms light travel times back and forth between
  Chicago and New York are indicated. The $y$-axis values are
  normalized to a per price-changing trade basis.}
\label{fig:dPdt}
\end{figure*}

When a 1-ms interval that ends with a price-\emph{increasing} in-force
E-mini trade has been identified, we search for near-coincident SPY
trades in each of the thirty 1-ms intervals prior to and following the
E-mini price-changing trade event. If a price-changing SPY trade has
occurred, we add the observed change in the SPY traded price
($\delta_l$) to an array that maintains a cumulative sum of these
price changes from ${\rm 
  lag}=-30~{\rm ms}$ to ${\rm lag}=+30~{\rm ms}$.
The foregoing procedure is also followed for price-{\it decreasing}
in-force E-mini trades. In the case of these declines, however, we add
$-1\times\delta_l$ to the array that maintains the cumulative
sums. This facilitates the combination of both price increases and
price decreases into a single estimator. Our composite response is
based on a total of $N$=14,078,656 price-changing E-mini trades
observed during the 577-day interval covered by the data and is
depicted in Figure \ref{fig:dPdt}, where we plot $dP/dt$ for SPY
vs. millisecond lag relative to E-mini price-changing trades. It is
clear that a larger SPY price response occurs at positive lag,
reinforcing the conclusion that price formation occurs primarily, but
not completely, at the CME.

\begin{figure*}[ht]
  \centering
  \includegraphics[width=0.9\textwidth,angle=0]{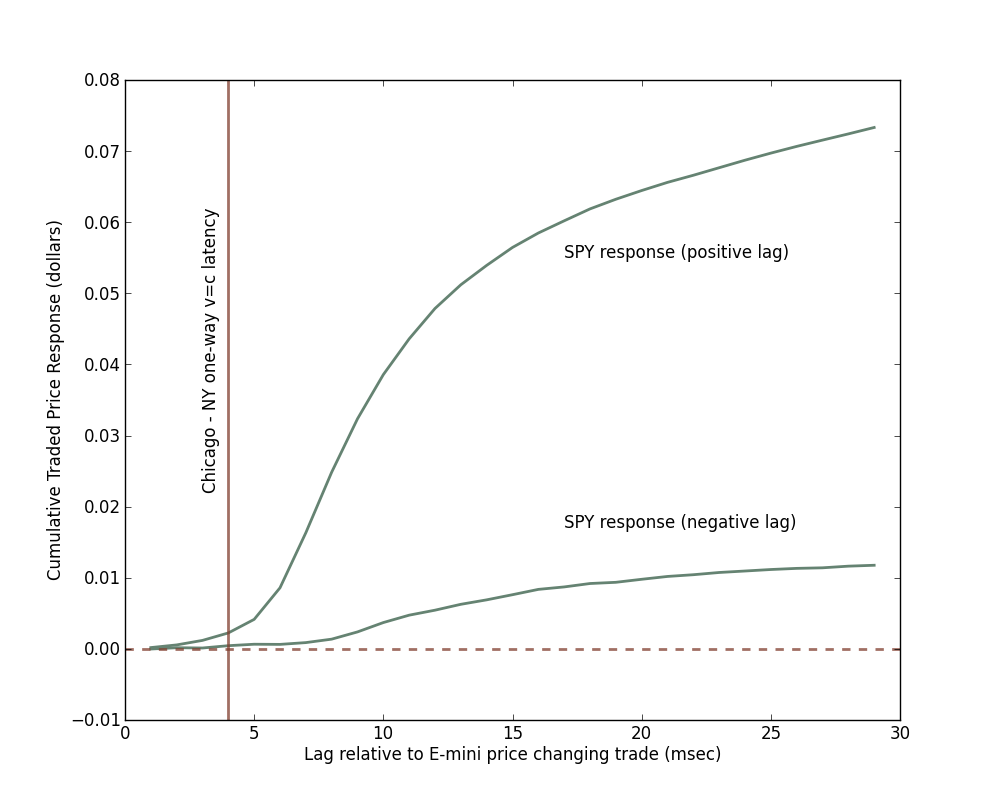}
  \caption{Cumulative Traded SPY price response to E-mini
    price-changing trades. The integrated SPY response over positive
    lags is a factor of 6.26$\times$ larger than the integrated SPY
    response over negative lags.}
\label{fig:Cumul1}
\end{figure*}

A cumulative {\it positive-lagged response}, $\Delta P(t_f)^{+}$, can
be quantified by integrating $dP/dt$ from 0 through lag $t_f$,
$$
\Delta P(t_f)^{+}= \int_{0}^{t_f} {dP\over{dt}}dt\, ,
$$ 
and a cumulative {\it negative-lagged response}, $\Delta P(t_f)^{-}$,
can be quantified by integrating $dP/dt$ from 0 through lag $-t_f$,
$$
\Delta P(t_f)^{-}= -\int_{-t_f}^{0} {dP\over{dt}}dt\, .
$$ 
$\Delta P(t_f)^{+}$ and $\Delta P(t_f)^{-}$ are plotted in Figure
\ref{fig:Cumul1}. It is clear that $\Delta P(t_f)^{+}>\Delta
P(t_f)^{-}$, again indicating that the E-mini is predominantly leading
SPY. The ratio ${\mathcal P}=\Delta P(t_f)^{+}/\Delta P(t_f)^{-}=6.26$
suggests that the E-mini is a factor of ${\mathcal P}\sim6.26$ times
more important than SPY in determining price formation. Trading on the
equity exchanges thus has a measurable impact, and price discovery is
not completely localized at the CME.

The sharing of price discovery between exchanges in Chicago and New
Jersey and the physical separation of these locations by roughly 1200
km means that the theoretical minimum round-trip time of information
flow is about 8 ms. This serves as an effective bound on inter-trade
duration, at which point trade rate will saturate. According to the
compound TMSMD model, with $\bar{k}=5$, an 8 ms inter-trade duration
corresponds to an annualized S\&P 500 volatility of approximately 36
percent. With $\bar{k}\leq4$, the corresponding model-implied
volatility is 45 percent or higher.

If underlying market pressure induces the trade rate to push against
the current distance-mediated limit, the resulting volatility could
lead to a large market move, either up or down. Prior to the SEC
imposition of coordinated limit rules among exchanges in April
of 2013, such a large market movement could have resulted in an E-mini
trading halt at the CME and a subsequent shift in trading
interest and price formation to equities exchanges in New Jersey. In
such a scenario, the physical separation ($d \leq 56$ km) among
exchanges in suburban New Jersey would correspond to a new effective
inter-trade duration bound of 0.3 ms or less, which our model suggests
is associated with a market volatility substantially greater than 100
percent. Fortunately, as of 8 April, 2013, the SEC has enforced a
coordination among circuit breakers that largely precludes such a
cascade of events. However, if some other market event resulted in
price formation occurring at a single physical location (either
Illinois or New Jersey), a period of market stress could lead to an
extremely high saturation point for asset trade rate and
correspondingly high market volatility.

The upshot is that coordinated limit rules between equities and
futures markets may be essential to prevent extreme volatility and
potential market failure. While the E-mini and SPY contracts do not
represent the entire market, they comprise a substantial fraction of
market liquidity and are indicative of what may happen among other
assets.

Our model suggests that physical separation of and distributed price
formation among exchanges has the benefit of imposing a market-wide
volatility ceiling. Our model also suggests that coordinated limit
rules among exchanges may be essential for maintaining market
stability. In fairness, the model can only be used to extrapolate
behavior during quiescent regimes (and not periods that are affected
by major news arrivals), when trade-time returns evolve as Gaussian
random walks. However, we believe the foregoing results highlight a
potential source of market weakness.

\section{Conclusion}
\label{conclusion}

The empirical and theoretical work of this paper is intimately linked
to the work of \citet{Mandelbrot1963}, \citet{Clark1973},
\citet{Brada1966}, \citet{Mandelbrot1967} and \citet{Ane2000}, which
show that fat-tailed returns distributions are consistent with a
Gaussian random walk subordinated by an appropriate stochastic
process. Our empirical insight is that after controlling for
pre-scheduled, market-wide news announcements, the subordinating
process for a highly liquid market aggregate (the E-Mini S\&P 500
near-month futures contract) is simply characterized by a model
of high-frequency trade arrival. Our theoretical contribution is to
develop a parsimonious inter-trade duration model that serves as an
appropriate subordinator, and to compose it with a Gaussian random
walk to arrive at a hierarchical model of returns in clock
time. Specifically, we have found that a modification of the
Markov-Switching Multifractal Duration model, developed by
\citet{Chen2013}, can be tuned to provide a statistically correct
description of the trade arrival process. Returns in the tail of the
distribution arise as a consequence of the faster random walks
generated by periods where the trading rate is high and volatility
persistence is generated by correlation among latent Markov-switching
shocks. The upshot is that outside of pre-scheduled news-affected
periods, the observed non-Gaussianity in the E-mini returns
distribution can be fully attributed to the temporal clustering of
trades.

We use our model to extrapolate the relationship between inter-trade
duration and market volatility and to explore conditions under which
market pressure could lead to catastrophically high volatility and
potential market failure. We note that this could occur during a
period of large market movements and under circumstances that cause
price formation to occur at a single physical venue, leading to a
reduction in the current distance-mediated effective bound on systemic
inter-trade duration. While our model is not tuned to deal explicitly
with regimes of market stress, our results may point to a potential
for weakness in the current system.

Further work appears warranted. The E-mini attracts a substantial
depth of book and routinely exhibits notional trading volumes in
excess of 100 billion dollars per day. As such, minor news events,
such as those related to a single company, will rarely lead to
tangible changes in the index. It is possible, however, that the
Gaussian spectrum of tick-time returns is a feature that is largely
specific to the heavily traded E-mini. Expanding the work of this
paper to a broader set of assets could lead to important innovations
to the model. Additionally, it would be useful to adapt the model to
explain the evolution of returns under conditions of market stress.

\newpage
\appendix
\section{Expectation of the Maximum of Exponential Random Variables}
\label{extremum}

Suppose that random variables $\{X_i\}_{i=1}^n$ are independently and
identically distributed as an Exponential with mean parameter
$\nu$. Then their common PDF and CDF are
\begin{align*}
  f_{X}(x) & = \frac{1}{\nu} \exp{\left(-\frac{x}{\nu}\right)} \\ 
  F_{Xi}(x) & = 1 - \exp{\left(-\frac{x}{\nu}\right)}. \\
\end{align*}
If we let $Y_n = \max\{X_1,X_2,\ldots,X_n\}$, then the CDF of $Y_n$ is
\begin{align}
  G_n(y) & = P(Y_n \leq y) \nonumber \\
  & = P(X_1 \leq y, X_2 \leq y, \ldots, X_n \leq y) \nonumber \\
  & = P(X_1 \leq y)P(X_2 \leq y) \cdots P(X_n \leq y) \nonumber \\
  & = \prod_{i=1}^n F_{X_i}(y) \nonumber \\
  & = F_{X}(y)^n \nonumber \\
  & = \left(1 - \exp{\left(-\frac{y}{\nu}\right)}\right)^n. \label{extremeCDF}
\end{align}
It follows that the PDF of $Y_n$ is
\begin{align}
  g_n(y) & = n F_X(y)^{n-1} f_X(y) \nonumber \\
  & = \frac{n}{\nu}
  \exp{\left(-\frac{y}{\nu}\right)}\left(1-\exp{\left(-\frac{y}{\nu}\right)}\right)^{n-1}, \label{extremePDF}
\end{align}
which can be shown to have expectation,
\begin{align}
  \E[Y_n|\nu] & = \int_{0}^{\infty} y
  \exp{\left(-\frac{y}{\nu}\right)}\left(1-\exp{\left(-\frac{y}{\nu}\right)}\right)^{n-1}
  dy \nonumber \\
  & = \nu \sum_{i=1}^n \frac{1}{i}. \label{extremeExp}
\end{align}


\newpage

\renewcommand{\baselinestretch}{1}
\selectfont
\nocite{*}
\bibliographystyle{asa}
\bibliography{tmsmd}

\end{document}